\documentclass[journal=aamick,manuscript=article]{achemso}
\usepackage{graphicx,psfrag}
\usepackage{makecell}
\usepackage[colorlinks=true,linkcolor=blue]{hyperref}
\usepackage{latexsym,mathrsfs,mathrsfs}
\usepackage{braket}
\usepackage{graphicx}
\usepackage{caption}
\usepackage{wrapfig} 
\usepackage[font=footnotesize,labelfont=bf,
justification=justified,
format=plain]{caption}
\usepackage{array}
\usepackage{lipsum} 
\usepackage{varioref}
\usepackage[utf8]{inputenc}
\usepackage{hyperref}
\hypersetup{colorlinks,%
	citecolor=blue,%
	filecolor=blue,%
	linkcolor=blue,%
	urlcolor=blue}
\usepackage{cleveref}

\author{Christina E. Antony}
\affiliation{School of Physics, Indian Institute of Science Education and Research Thiruvanthapuram, Vithura, Thiruvananthapuram, India-695551}
\author{Praveen S. G.}
\affiliation{School of Physics, Indian Institute of Science Education and Research Thiruvanthapuram, Vithura, Thiruvananthapuram, India-695551}
\author{Adithya Jayakumar}
\affiliation{School of Physics, Indian Institute of Science Education and Research Thiruvanthapuram, Vithura, Thiruvananthapuram, India-695551} 
\author{Gaana K.}
\affiliation{School of Physics, Indian Institute of Science Education and Research Thiruvanthapuram, Vithura, Thiruvananthapuram, India-695551} 
\author{Akshay Yadav}
\affiliation{School of Physics, Indian Institute of Science Education and Research Thiruvanthapuram, Vithura, Thiruvananthapuram, India-695551}
\author{Nikhil S. Sivakumar}
\affiliation{Department of Physics, IISER Mohali, Knowledge city, Mohali-140306, India}
\author{Niranjan Kamath}
\affiliation{Department of Physics, IIT Madras, Chennai-600036, India}
\author{Suma M. N.}
\affiliation{LPSC Bangalore, ISRO, Benguluru-560008, India}
\author{Vinayak B. Kamble}
\affiliation{School of Physics, Indian Institute of Science Education and Research Thiruvanthapuram, Vithura, Thiruvananthapuram, India-695551}
\author{D. Jaiswal-Nagar}
\email{deepshikha@iisertvm.ac.in}
\affiliation{School of Physics, Indian Institute of Science Education and Research Thiruvanthapuram, Vithura, Thiruvananthapuram, India-695551}

\title[An \textsf{achemso}]
{Metal-polymer hybrid chemiresistive sensor for low concentration fast hydrogen detection}


\begin{document}

%\correspondingauthor{\textsuperscript{2}To whom correspondence should be addressed. E-mail: deepshikha@iisertvm.ac.in}

%\keywords{Hydrogen Sensors $|$ Chemiresistive sensors $|$ Palladium metal $|$ X-ray photoelectron spectroscopy} 
\begin{abstract}
Low concentration hydrogen gas detection is of paramount importance both in space applications as well as medical applications. It is also critically important for safe handling of hydrogen below the explosive limit. Here, we report a novel hybrid Pd metal-polymer chemiresistive sensor that can sense 0.5$\%$ hydrogen (H$_2$) gas in ambient conditions of temperature and pressure with the highest reported sensitivity ($\sim$ 30$\%$) obtained earlier by a physical deposition technique, making it an extremely good sensor for real life low concentration hydrogen gas detection. The sensor is easy to fabricate and is also extremely cost-effective for commercial applications. The obtained hybrid chemiresistive sensor comprises palladium (Pd) nanocrystals bound by oxygen and nitrogen atoms of a stabilizer Polyvinylepyrollidone (PVP), grown on top of a self-assembled monolayer. The exceptional rise time-constant is proposed to arise from hydrogen loading at the (111) surface of the palladium nanocrystal which is a very fast process and subsequent fast diffusion of the H atoms from the surface into the bulk. An effort to increase the number of available sites by UV-ozone cleaning, resulted in a degradation of the sensing device due to the poisoning of the available sites by oxygen. 
\end{abstract}
%\date{\today}
\maketitle
Hydrogen gas is an efficient, renewable and clean energy source that can contribute to solving the problems of depleting world fossil fuel reserves \cite{yurum}. However, hydrogen is a hazardous gas since it is flammable and explosive in air and the explosive limit ranges between 4.65 and 93.9 vol$\%$ hydrogen (H$_2$). Since it is colourless, odourless and tasteless, it is difficult to detect H$_2$ by human senses. Therefore, alternate robust hydrogen leakage sensors are required to rapidly sense and quantify the concentration of hydrogen much below the regulatory range to avoid any potential explosion hazard associated with using hydrogen as an energy source \cite{christofides}. Hydrogen sensors are also required in the chemical industry to monitor hydration of hydrocarbons, synthesis of methanol and ammonia, desulphurisation of petroleum products etc. \cite{hubert}. Breath hydrogen is an important parameter used in clinical determination of lactose intolerance, bacterial growth, microbial activity, diabetic gastroparesis etc. \cite{rizkalla} making hydrogen sensors that could detect hydrogen in the low concentration of 1000 ppm, of immense importance in biomedical applications. Hydrogen sensors are also needed in aerospace applications, especially those involving hydrogen fuel propulsion systems, where detection of hydrogen at low concentrations is necessary to monitor shuttle launches in order to avoid possible launch failures \cite{christofides,hunter}. Similarly, low concentration hydrogen detection helps in space exploration to estimate details of universe formation and its evolution \cite{krall}. So, it is of paramount importance to develop a safe, reliable, compact, cost-effective and efficient hydrogen sensor that can not only measure hydrogen gas in low concentrations with high sensitivity and fast response times but is also easy to fabricate.\\
Nanostructured materials, whose properties vary significantly compared to their bulk counterparts due to large surface to volume ratios and quantum-confinement effects, have been at the forefront of hydrogen sensing research \cite{zeng,xu,lee,lee1,kumar,ramanathan,ibanez,shin,wang,jeon,huang}.  A variety of hydrogen sensors in the form of nanofibres \cite{wang}, nanowires \cite{zeng}, nanoclusters \cite{shin} etc. have been reported, that utilize a change in some physical property like mass, volume, optical constant, work function, resistivity \cite{zeng,xu,lee,lee1,kumar,ramanathan,ibanez,wang,jeon,huang,shin} etc.  Of these, resistivity based electronic sensors offer an appealing possibility of cost-effectiveness, durability and reliability due to the ease of operation, integration with conventional circuit electronics, miniaturization potential, scalability and portability. Palladium (Pd) metal’s hydrogen selective catalytic activity is very well known \cite{noh,xu,kay,lewis,flanagan}. In this process, the surface of palladium acts catalytically on the incoming hydrogen molecule breaking the diatomic hydrogen to monoatomic hydrogen which then diffuse inside the Pd lattice \cite{kay} and occupy interstitial sites of Pd resulting in the expansion of Pd lattice and formation of a hydride of palladium, PdH$_x$ \cite{noh,xu,kay,lewis,flanagan}. For x $<$ 0.01, a solid solution, PdH$_{\alpha}$, gets stabilised due to random occupation of hydrogen while for 0.02 $<$ x $<$ 0.6, an octahedral lattice hydride PdH$_{\beta}$ is formed \cite{flanagan,lewis}. The expansion is smaller than 0.13$\%$ in the PdH$_{\alpha}$ state (a$_0$ = 0.3890 nm for pure Pd and a$_{\alpha},_{max}$ = 0.3895 nm for PdH$_{\alpha},_{max}$) but reaches a change of 3.5$\%$ from PdH$_{\alpha},_{max}$ to PdH$_{\beta},_{min}$. So, bulk Pd based sensors suffer from a lower limit of sensing since the PdH$_{\alpha}$ to PdH$_{\beta}$ transition happens at $\sim$ 1$\%$ of H$_2$ concentration. However, the situation changes drastically when Pd is in a nanostructured form since the surface to volume ratio increases manifolds implying large surface activities, and hence, reducing the sensing limit to lower values \cite{dharmendra,xu}.\\ 
Chemiresistive sensors are a class of sensors that work on the change in electrical resistance due to surface adsorption of analyte molecules \cite{favier}. Such sensors are easy to fabricate, are very cost effective and have high miniaturization potential \cite{ibanez}. In this work, we report on a novel chemiresistive sensor that was built by capping Pd nanocrystals with a stabilizer Polyvinylepyrollidone (PVP). It is known that capping ligands chemisorbed on the surfaces of metal nanocrystals not only act as stabilisers but also modify the functionality of the nanocrystals since the attached atoms or functional groups of the stabilisers affect the nanoparticle’s accessible surface area, electronic structure properties and surface properties like catalysis \cite{collins,garcia,grammatikopoulos,teranishi}. In this regard, even though Pd-PVP structures have been investigated immensely for their catalytic properties \cite{garcia,collins,xian,teranishi}, there have been, surprisingly very few reports on its H$_2$ sensing abilities.  Here, we make a novel architecture of Pd-PVP structure that was built on a siloxane self- assembled monolayer. The PVP-Pd mole ratio was varied from 0.06 to 0.24 and each architecture that was built for a given PVP-Pd ratio, was tested for hydrogen sensing. It was found that only the assembly with PVP-Pd ratio of 0.10 could sense hydrogen. The sensing characteristics were found to be very good, wherein, the assembly could sense low concentrations of H$_2$ (upto 0.1$\%$) with fast response times ($\sim$ 50 s) and very high sensitivity (~ 30$\%$ at 0.5$\%$ H$_2$). From detailed x-ray photoelectron spectroscopy measurements, it was found that the chemisorption of the PVP molecule with the Pd nanocrystals happened via the oxygen and nitrogen atoms of PVP molecule. It was also found that the number of bare Pd metal active sites are maximum for PVP-Pd mole ratio of 0.10, resulting in a sensing response for only this ratio. The obtained chemiresistive sensor has immense potential for commercial applications due to its simple and cost-effective method of fabrication that yielded comparative results on sensitivity and response time obtained by a physical deposition method \cite{xu}.
\section*{Materials and Methods}
\subsection*{Materials:}
Palladium Chloride (PdCl$_2$) which is the metal precursor, was purchased from Merck-Aldrich and used as such. Poly (N-vinyl-2-pyrrolidone) (PVP, standard molecular weight of 40,000; Merck-Aldrich) was used as a stabilizer for the Palladium nanocrystals. Ethanol (C$_2$H$_5$OH) used as the reducing alcohol, was of guaranteed grade and used without any further purification. Siloxane (N- chloro(dimethyl)octyl silane, Merck-Aldrich) used for making a self-assembled monolayer (SAM) was used without any further purification. Synthetic air and hydrogen gas (4$\%$ dilution) cylinders were purchased from Bhuruka Gas, India. 
\subsection*{Palladium nanocrystals synthesis:}
The synthesis procedure comprises reduction \cite{teranishi} of metal precursor using ethanol in the presence of capping agent, PVP, at an elevated temperature. So, an aqueous solution (2mM) of H$_2$PdCl$_4$ was prepared by mixing 106.4 mg of PdCl$_2$ (0.6 mmoL)- the metal precursor, 6 mL of 0.2 M HCl and 294 mL of distilled water. By mixing 15 mL of as prepared 2mM H$_2$PdCl$_4$ solution with varying concentrations of ethanol (ranging from 10$\%$ to 70$\%$ vol.) and designated amount of PVP (ranging from 0.2 mg to 0.5 mg), different solutions were made. We made four different solutions with PVP$/$Pd ratio as 0.06, 0.10, 0.13 and 0.24. After letting the colloidal nanoparticles to age for two to three days, the solutions were refluxed in a 250 mL flask for about 15 hours at an elevated temperature of 378 K, as shown in Fig. \ref{fig:synthesisschematic}. Refluxing results in repeated filtration of the as prepared colloidal solution, resulting in PVP protected palladium (PVP-Pd) nanocrystals of a narrow size distribution see (Figs. \ref{fig:TEM} (a)-(d)).
\subsection*{Sensing assembly fabrication:}
To prepare the substrates for hydrogen gas sensing, glass slides were cut into rectangles of 1.0 cm x 0.3 cm sizes and degreased by sonicating in an ultrasonic bath using acetone and isopropanol, for sufficient  amount of time. The cleaned slides were, then, washed with milliQ water. The centre of the cleaned glass slides was shadow masked with a copper wire of 110 $\mu$m diameter. A thin layer of gold (Au) film of thickness 50 nm was deposited on the glass slide using a thermal evporator, to act as contact pads. Prior to gold layer deposition, a thin wetting layer of chromium was also deposited. Then, the shadow mask was removed to give an insulating channel of 80 $\mu$m width. The fabricated substrate was, then, immersed in 0.1M chloro (dimethyl) octyl silane for 1 hr which results in the formation of a silane self-assembled monolayer (SAM). In the final stage of sensing assembly fabrication, colloidal PVP-Pd nanocrystal solution, prepared according to the details above, was drop casted (20$\mu$l) over the channel in the silane SAM coated substrate. The assembly was, then, allowed to dry in the room temperature conditions. The detailed schematic of the entire procedure is shown in Fig. \ref{fig:synthesisschematic}.
\subsection*{Sensing measurement:}
Resistance (R) based hydrogen (H$_2$) gas sensing measurements were carried out in an in-house built gas sensor characterization system \cite{suresh}. H$_2$ and synthetic air flows towards the chamber were regulated using Alicat Scientific’s mass flow controllers to achieve desired dilution of the H$_2$ gas. Prior to hydrogen gas flow, the chamber was purged and saturated with synthetic air. The change of resistance of the devised assembly described above, upon the sorption of hydrogen was measured using Keithley’s 2700 Digital multimeter cum data acquisition system \cite{suresh,dharmendra}.
\subsection*{Experimental techniques:}
Ultraviolet-visible (UV-Vis) spectroscopy was done on Shimadzu's UV-VIS-NIR spectrophotometer (Model No. UV 3600). The surface morphology  and the size distribution of the as-formed PVP-Pd nanocrystals were characterized using a 300 kV FEI Tecnai’s (Model No. G2 F30 S – Twin) high resolution transmission electron microscope (HRTEM). Samples for the HRTEM measurements were prepared by dropcasting 20$\mu$l of the colloidal solution onto a TEM grid, followed by natural evaporation of the solvent. To understand the final chemical and electronic state of PVP-Pd assembly, high resolution x-ray photoelectron spectroscopy experiments were done using a ESCA Plus spectrometer (Omicron Nanotechnology Ltd. Germany) equipment with Mg K$_{\alpha}$ radiation (1253.6 eV). The instrument is equipped with an auto-charge neutraliser.  
\section*{Metal-Polymer chemiresistive hydrogen sensor architechture}
\begin{figure}
	\centering
	\includegraphics[width=1\linewidth]{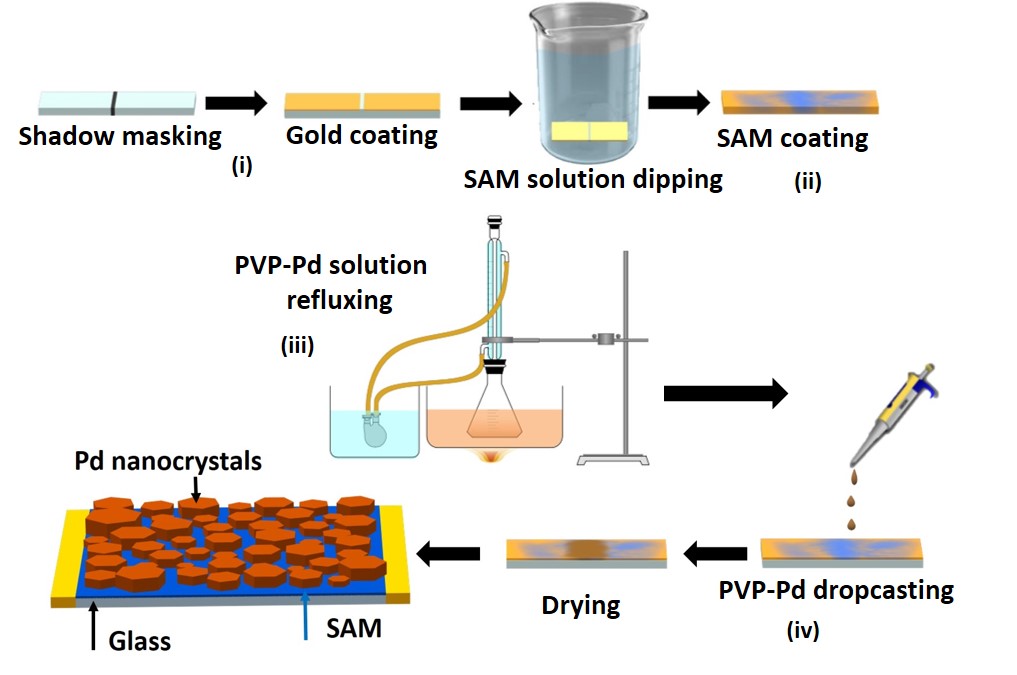}
	\caption{Schematic of the steps involved in making the chemiresistive sensor.}
	\label{fig:synthesisschematic}
\end{figure}
As described in the methods section, the novel architechture of our hybrid metal-polymer chemiresistive sensor consists of a random array of hexagon shaped Pd metal nanocrystals complexed with PVP molecules, on a siloxane self-assembled monolayer (SAM) coated on top of bare glass as shown in Fig. \ref{fig:synthesisschematic}. The whole process is divided in four steps (i) fabrication of the substrate with a channel region in which the palladium nanocrystals can grow, (ii) growth of self-assembled monolayer (SAM) in the channel region where the particular SAM was chosen for its known ability of reducing stiction on the surface of bare glass and enhancing the sensitivity of hydrogen response \cite{xu}, (iii) preparation of the colloidal PVP-Pd nanocrystal solution by using the refluxing technique and (iv) drop-casting the colloidal PVP-Pd nanocrystal solution over the channel in the silane SAM coated substrate. In order to find the optimal PVP-Pd hybrid complex that could give a resistive sensing response, different solutions made of varying PVP$/$Pd mole ratios from 0.24 to 0.06, were drop-casted onto the SAM layer. The final assembly is shown in Fig. \ref{fig:synthesisschematic}. 

\begin{figure}
	\centering
	\includegraphics[width=1\linewidth]{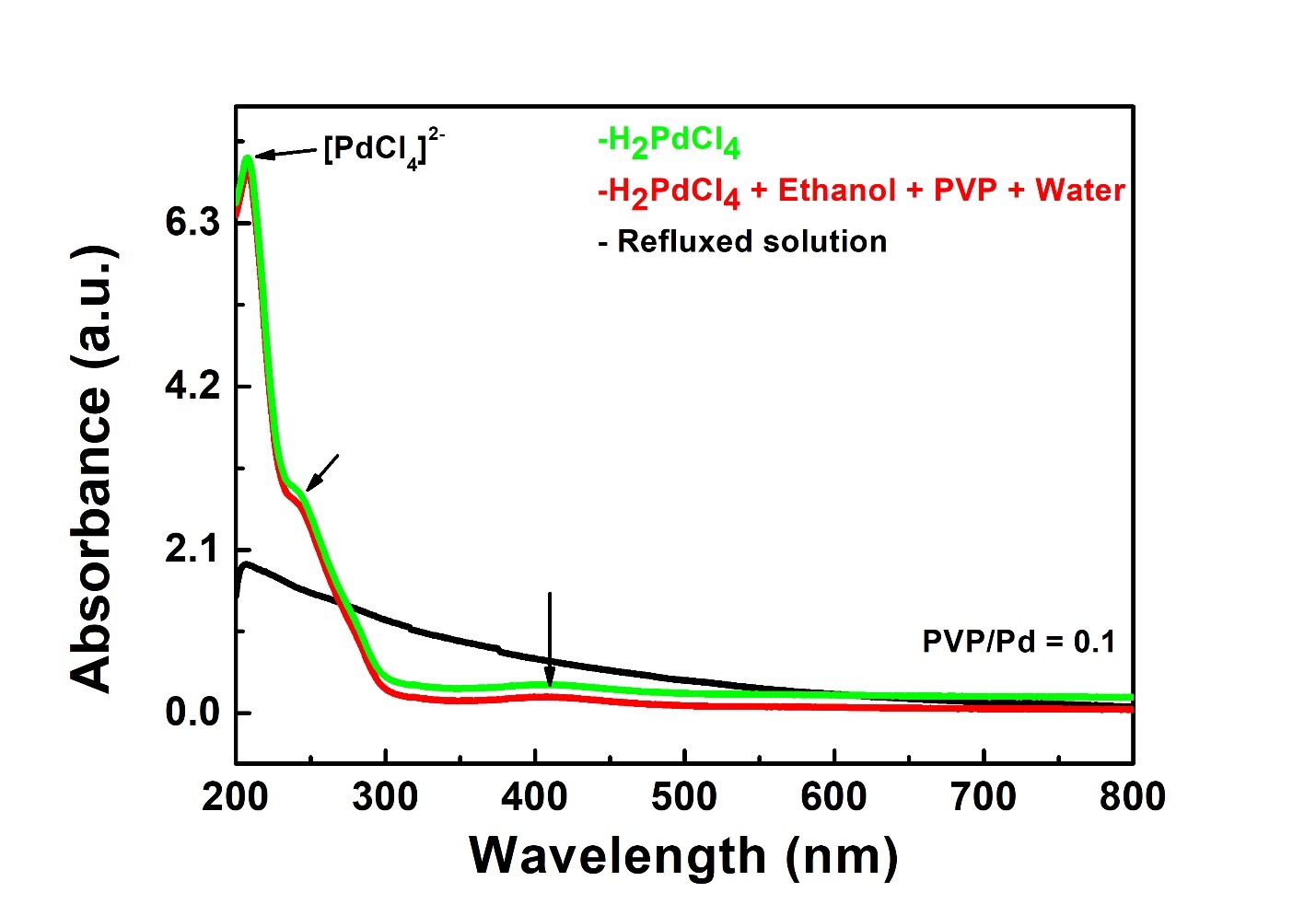}
	\caption{(Colour online): UV-VIS spectra obtained for H$_2$PdCl$_4$ (green), unrefluxed solution of H$_2$PdCl$_4$ together with ethanol, water and PVP mixed with Pd in a PVP/Pd mole ratio of 0.1 (red) and the corresponding refluxed solution (red).}
	\label{fig:UV-Vis}
\end{figure}

The H$_2$PdCl$_4$ solution with ethanol, PVP and water is pale yellow in colour without any refluxing, due to the formation of [PdCl$_4$]$^{2-}$ (Pd(II)) ions \cite{teranishi1,sil,berger,teranishi}. These Pd(II) ions form stable complex with nitrogen and oxygen containing organic compounds, so they co-ordinate to PVP molecules before the reduction reaction \cite{berger,teranishi1}, resulting in a PVP-Pd(II) complex. This complex results in  peaks at $\sim$ 220 nm, 240 nm and 410 nm in the UV-VIS spectrum \cite{teranishi,teranishi1,sil}, as evident in the spectra marked by arrows in Fig. \ref{fig:UV-Vis}. Green solid line in Fig. \ref{fig:UV-Vis} represent the spectra taken with only the H$_2$PdCl$_4$ solution while the red solid line corresponds to the spectra with ethanol, PVP and water added to H$_2$PdCl$_4$ and allowed to wait without any refluxing. As can be clearly seen, the difference in the two spectra (H$_2$PdCl$_4$ alone and H$_2$PdCl$_4$ mixed with PVP, alcohol and water) is negligible. This implies that without refluxing, [PdCl$_4$]$^{2-}$ ions remain in the solution and Pd(0) atoms are not formed. However, when the solution is refluxed, all the three peaks disappear and the absorption spectrum exhibited a typical light-scattering phenomenon, as evidenced in the spectra displayed in black colour in Fig. \ref{fig:UV-Vis}, clearly suggesting that [PdCl$_4$]$^{2-}$ ions got reduced from Pd(II) state to Pd(0) state. During  refluxing, Pd(0) ions are stabilized by the hydrophobic interactions between the hydrophobic alkyl groups of PVP and the metallic Pd surface \cite{kockzur}. Additionally, the absorption in the visible region increased compared to the ultra-violet region, suggesting the formation of band-like structure for Pd nanocrystals.

\begin{figure}
	\centering
	\includegraphics[width=1\linewidth]{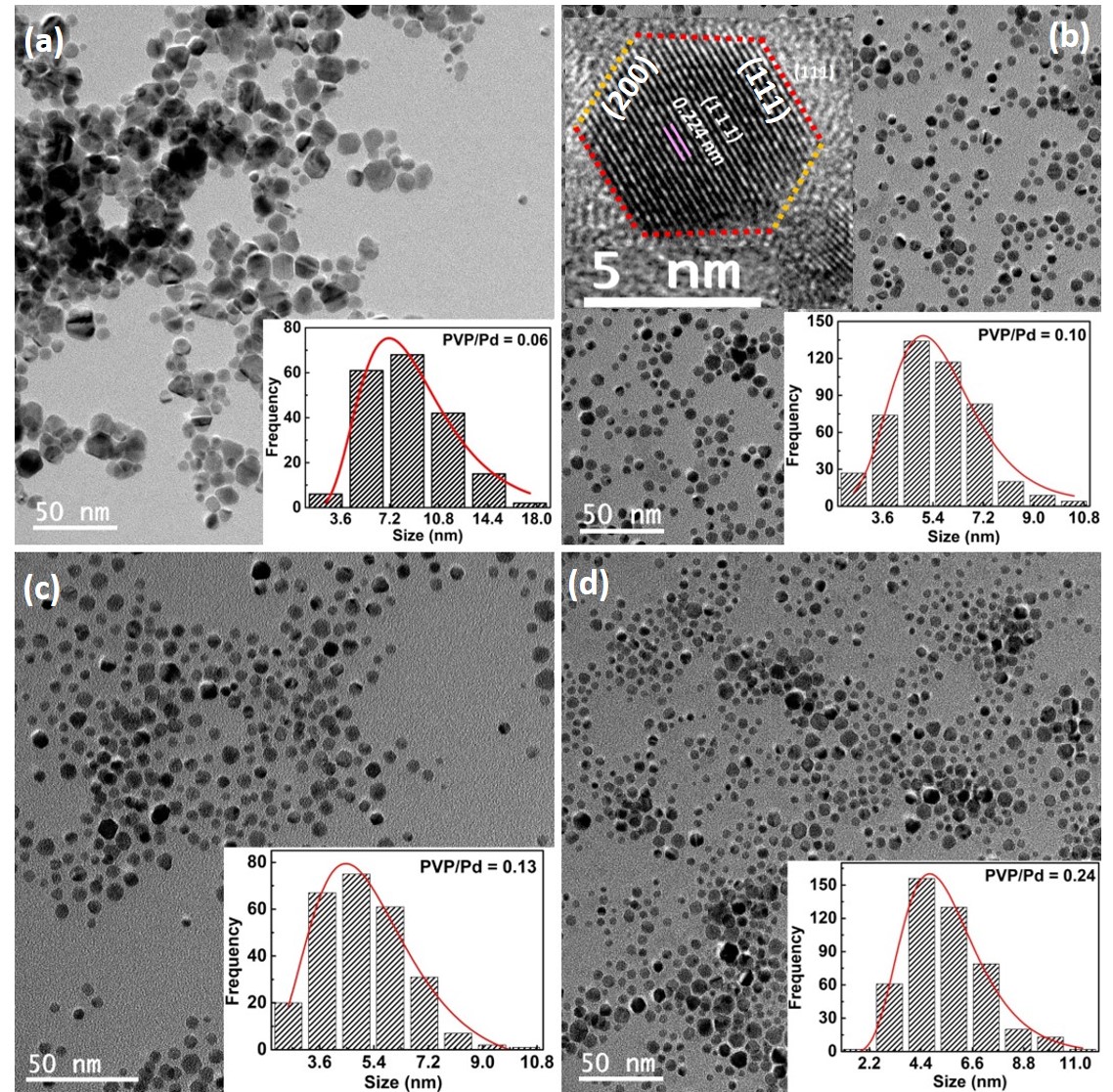}
	\caption{TEM micrographs of PVP-Pd system obtained for PVP/Pd mole ratio of (a) 0.06, (b) 0.10, (c) 0.13 and (d) 0.24. Inset of each shows a histogram of the sizes of the resultant Pd NC calculated using the Image J and Scanning Probe Image Processor software both of which gave consistent results. Average size obtained by fitting a log-normal distribution to the histograms was obtained as 8.5 nm, 5.5 nm, 5.1 nm, and 5.3 nm for PVP/Pd ratio of 0.06, 0.1, 0.13 and 0.24 respectively. Top left inset of (c) shows a HRTEM image of a  six faceted nanocrystal obtained for PVP/Pd mole ratio of 0.10. Obtained d-spacing for (111) and (200) facets have been indicated.}
	\label{fig:TEM}
\end{figure}

Since the functionality of the hybrid PVP-Pd complex is assumed to be highly dependent on the size of the Pd nanocrystal (NC) \cite{xian}, wherein, a small size was found to have a higher catalytic activity, the hope is that a small size of the complex has a better possibility of it being responsive to hydrogen gas, and hence, behaving as a resistive hydrogen sensor. In order to check the average size of the grown Pd nanocrystals, we subjected them to TEM measurements, as shown in Fig. \ref{fig:TEM}. From the main panels of Figs. \ref{fig:TEM} (a)-(d), it was found that the PVP-Pd NC’s have a hexagonal shaped morphology with uniform and uni-modal distribution (see the bottom inset of each Fig. \ref{fig:TEM} (a)-(d)). Top left inset of Fig. \ref{fig:TEM} (b) shows a representative HRTEM image of a PVP-Pd NC produced for a PVP$/$Pd mole ratio of 0.10. All the NC’s produced for other mole ratio’s have a similar morphology (see Supplementary Fig. 1). The NC was captured with a (111) facet (shown as red dotted lines) and (200) facet (shown as yellow dotted lines) \cite{grammatikopoulos}. From the obtained average size of the hybrid PVP-Pd complex, it is found that varying the PVP$/$Pd ratio from 0.06 to 0.24 results in an average size of $\sim$ 5 nm for all conplexes barring the ones obtained for PVP$/$Pd ratio of 0.06 for which the size is slightly higher at 8.5 nm such that all the NC's grow with two facets, (111) and (200). So, a small size of $\sim$ 5 nm gives us a possibility of the hybrid PVP-Pd complex to be responsive to hydrogen gas.

\section*{Hydrogen sensing at ambient conditions:}
\begin{figure}[h]
	\centering
	\includegraphics[width=1\linewidth]{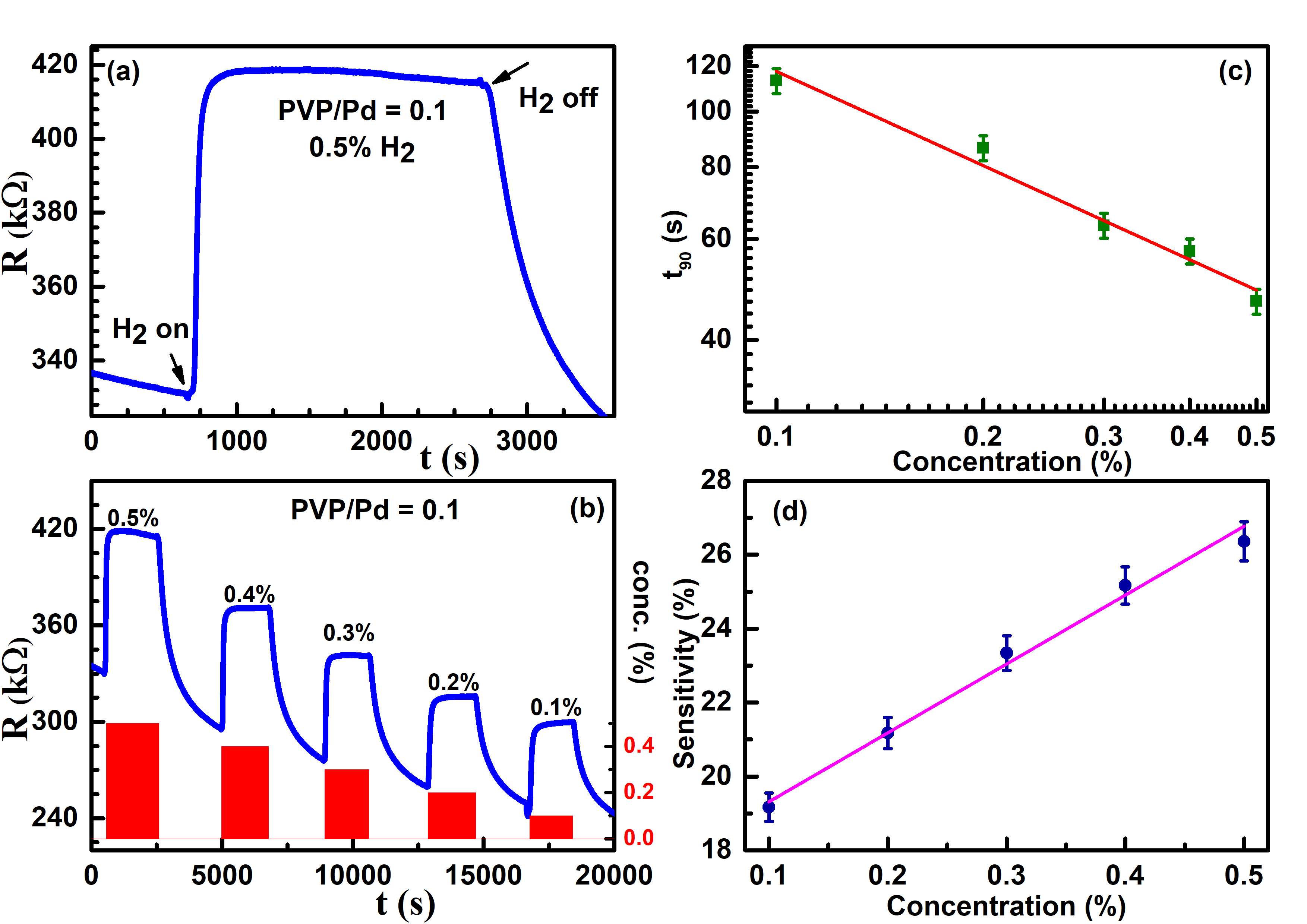}
	\caption{Hydrogen sensing in the architecture with hybrid PVP$/$Pd complex of mole ratio 0.10. (a) Increase in resistance upon exposure to 0.5$\%$ H$_2$ gas and subsequent saturation. Onset of H$_2$ loading and unloading are indicated by arrows. (b) Same architecture of (a) responsive to step-wise decrease of H$_2$ gas concentration from 0.5$\%$ upto 0.1$\%$ shown as red bars of progressively decreasing height. Red labels on the right indicate the concentration in percentage. (c) Log-log plot of the time-constant, $t_{90}$, with $\%$ H$_2$ gas concentration to identify the kind of diffusion process at work. Red solid line is a straight fit to the equation \ref{eqn:logt} and the obtained power-law exponent $r$ is 0.5. (d) Increase of sensitivity, as defined by equation \ref{eqn:sensitivity}, with percentage concentration. Pink solid line is a straight line fit to the data with the slope $m$ = 18.65 and intercept $c$ = 17.45.}
	\label{fig:hydrogensensing}
\end{figure}
To check the hydrogen sensing capabilities of the four different architectures made with varying PVP$/$Pd mole ratios (0.06, 0.1, 0.13 and 0.24) prepared as described above and having similar sizes, each of them were subjected to hydrogen gas of different concentrations. However, even after repeated trials with architectures prepared in different batches and hydrogen gas exposures of very high concentrations, it was found that only the architecture with PVP$/$Pd mole ratio of 0.10, responded to hydrogen gas. Fig. \ref{fig:hydrogensensing} (a) shows the room temperature resistive response to 0.5$\%$ H$_2$ gas concentration in the architecture with PVP$/$Pd mole ratio of 0.10 and a SAM concentration of 0.1 M. Prior to hydrogen exposure, synthetic air was flown in the sensing chamber to saturate the resistance of the device. As can be observed from Fig. \ref{fig:hydrogensensing} (a), a perfect saturation was not reached and a very slight slope is observed before hydrogen loading on the device at a starting resistance of $\sim$ 330 k$\Omega$. From  \ref{fig:hydrogensensing} (a), it is immediately clear that the device shows an extremely good response to hydrogen gas, where the resistance increases from 330 k$\Omega$ to 420 k$\Omega$ at which value, it saturates. The rise response time, $t_{90}$, defined as the time taken to reach 90$\%$ of the saturation value was found to be $\sim$ 47 s. By comparing the t$_{90}$ response of various Pd based resistive sensors under different concentrations of H$_2$ gas (see Table 1 of Supplementary), it can be said that our chemiresistive H$_2$ sensor shows an exctremely good rise time response in ambient H$_2$ conditions, corresponding to a real H$_2$ leak environment.\\  
To investigate the sensing performance of the architecture with hybrid PVP$/$Pd complex of mole ratio 0.10 under exposure to lower concentrations of H$_2$, we subjected the device to concentrations of 0.4$\%$, 0.3$\%$, 0.2$\%$ and 0.1$\%$ H$_2$ as shown in Fig. \ref{fig:hydrogensensing} (b). The response characteristic to hydrogen at lower concentrations is similar to that at 0.5$\%$, namely, a very fast increase in resistance upon hydrogen loading followed by a near-perfect saturation and a slower decrease on hydrogen unloading. Gardner et al. \cite{gardner} considered a gas diffusion model in a porous thin film of thickness $x_0$, wherein, the gas diffuses inside the thin film with diffusivity $D$ of the free gas (related to the film porosity and temperature) and immobolises at free surface sites. They found the steady-state diffusion time-constant to vary with concentration $C$ as:
\begin{equation}\label{eqn:time-constant}
\tau = kx_0^2\frac{C^{r-1}}{D}
\end{equation}
where $k$ is related to adsorption/desorption rate constant and $r$ is the power-law exponent. So, for a given film thickness $x_0$ and diffusivity $D$, the above equation can be rewritten as
\begin{equation}\label{eqn:logt}
log\tau = A + (r-1)logC;~~~~~~~~~~~      A = log\bigg(\frac{kx_0^2}{D}\bigg)
\end{equation}
The value of the power-law exponent $r$, then, defines the diffusion process: for $r >$ 1, the diffusion process is said to be \textit{quasi}-slow while for $r <$ 1, the diffusion process is assumed to be \textit{quasi}-fast. Lower the value of $r$, faster is the process \cite{gardner,yadav}. In order to understand the kind of difussion process that may be at play in our novel architecture, a log-log plot of $t_{90}$ vs. concentration $C$ was made as shown in Fig. \ref{fig:hydrogensensing} (c). Since a $t_{90}$ time-constant mimicks the steady state time-constant $\tau$ very well, equation \ref{eqn:logt} can be applied to the experimentally obtained data points, as shown by the red solid line. A small value of $r$ = 0.5 obtained from the fit, then, points to the fact that the diffusion of H atoms from the surface into the Pd complexed PVP film is a very fast process. It was found \cite{johnson,zalineeva} that H adsorption on the (111) surfaces compared to those of (100) of a Pd nanocrystal results in a very fast H$_2$ loading. So, the fast diffusion process of H atoms implies that the H atoms initially adsorb preferentially on the (111) surfaces of the hybrid Pd-PVP nanocrystal rather than on its (200) surface \cite{johnson,zalineeva}.\\  
Time-constant and sensitivity are the two parameters that are usually used to define the performance of a sensing device \cite{gardner,jeon,johnson,zalineeva}. So, it is prudent to check if the very good time-constant values obtained in Figs. \ref{fig:hydrogensensing} (a)-(c) are also accompanied by good sensitivity values. Accordingly, Fig. \ref{fig:hydrogensensing} (d) shows the sensitivity variation with $\%$ H$_2$ concentration, defined in the usual way \cite{lee,noh} as:

\begin{equation}\label{eqn:sensitivity}
S = \frac{\Delta R}{R} ~~ \times~~ 100\% = \frac{R_{H_2}-R_{Air}}{R_{Air}}~~\times~~ 100\%
\end{equation}

where $R_{Air}$ is the resistance of the sensor exposed to air only
and $R_{H_2}$ represents the resistance after exposure to hydrogen gas. From the graph, it is immediately apparent that excellent values of sensitivity are obtained in our PVP-Pd novel architecture, where a sensitivity value of 26.7$\%$ is obtained at 0.5$\%$ H$_2$ concentration, similar to the one obtained on SAM covered glass films made with physical vapour deposition method \cite{xu} (see Table 1 of Supplementary). In a zinc oxide$/$indium oxide core–shell nanorod based hydrogen sensor prepared by sputtering, a sensitivity of 20.5 was obtained for 0.05 $\%$ of hydrogen \cite{huang}, which on extrapolation to 0.5$\%$ of hydrogen, doesn't make it go beyond $\sim$ 27$\%$.\\ 
In bulk Pd, resistivity increase follows the Sievert’s law where it varies as square root of hydrogen concentration:
\begin{equation}
Sensitivity = \frac{1}{K_s}(pH_2)^{1/2}
\label{eqn:sievert}
\end{equation}
where $K_s$ is the Sievert's constant and p$H_2$ is the H$_2$ partial pressure in the environment \cite{lewis,sieverts,mueller}. However, from Fig. \ref{fig:hydrogensensing} (d), it can be seen that a straight line fits the $Sensitivity$ vs. H$_2$ concentration data very well. A similar linear variation in sensitivity with H$_2$ gas concentration has also been observed in other reports \cite{ibanez,lee1,xu}. So, a stronger exponent of 1 compared to the 0.5 value expected from equation \ref{eqn:sievert}, suggests that the amount of hydrogen adsorbed on available sites of our novel architecture is much higher compared to bulk for a given concentration, due to the very small size ($\sim 5.5 nm$) of the nanocrystals \cite{zalineeva} as well as the existence of (111) and (200) facets in the nanocrystals \cite{johnson} increasing the surface to volume ratio of a nanocrystal compared to a bulk crystal manifolds. Consequently, the number of available Pd sites are quite large in our novel architecture.\\
H$_2$ gas sensing on bulk Pd happens by catalytic dissociation of the incoming H$_2$ gas molecules into two H atoms at the Pd surface, which then move inside the Pd lattice forming PdH$_x$. The H$_2$ dissociation reaction on the Pd surface is expected to happen as \cite{sun}:
\begin{equation}
H_2+2S_{Pd} \rightarrow 2H\mbox{-}S_{Pd}
\label{eqn:dissociation}
\end{equation}
where $S_{Pd}$ denotes the available surface sites on Pd nanocrystals.\\
In PdH$_x$, when x $<$ 0.01, a solid solution of PdH$_{\alpha}$ gets formed due to random occupation of hydrogen, while for 0.02 $<$ x $<$ 0.6, an octahedral lattice hydride PdH$_{\beta}$ is formed. So, in the low concentration range of  0.1$\%$ to 0.5$\%$ of our consideration, a solid solution of PdH$_{\alpha}$ is stabilised. Hence, the rise in resistance of these structures may, possibly, be happening due to increase in work function of the PVP-Pd complex due to surface adsorption of the dissociated H atoms at the available Pd sites \cite{dharmendra,barr,wu,morris,sun}. A linear increase in sensitivity with H$_2$ concentration, then, implies progressivly increasing work function due to progressively increasing surface adsorption of H atoms. This, in turn, implies that the dissolved H atoms inside the Pd lattice take up the conduction electrons of Pd and become hydrogen anions \cite{sakaguchi}. The loss of electrons from the conduction band of the Pd lattice lowers the Fermi level of the system resulting in an increase in the work-function of the system \cite{sakaguchi}.

\begin{figure}[tbhp]
	\centering
	\includegraphics[width=1.1\linewidth]{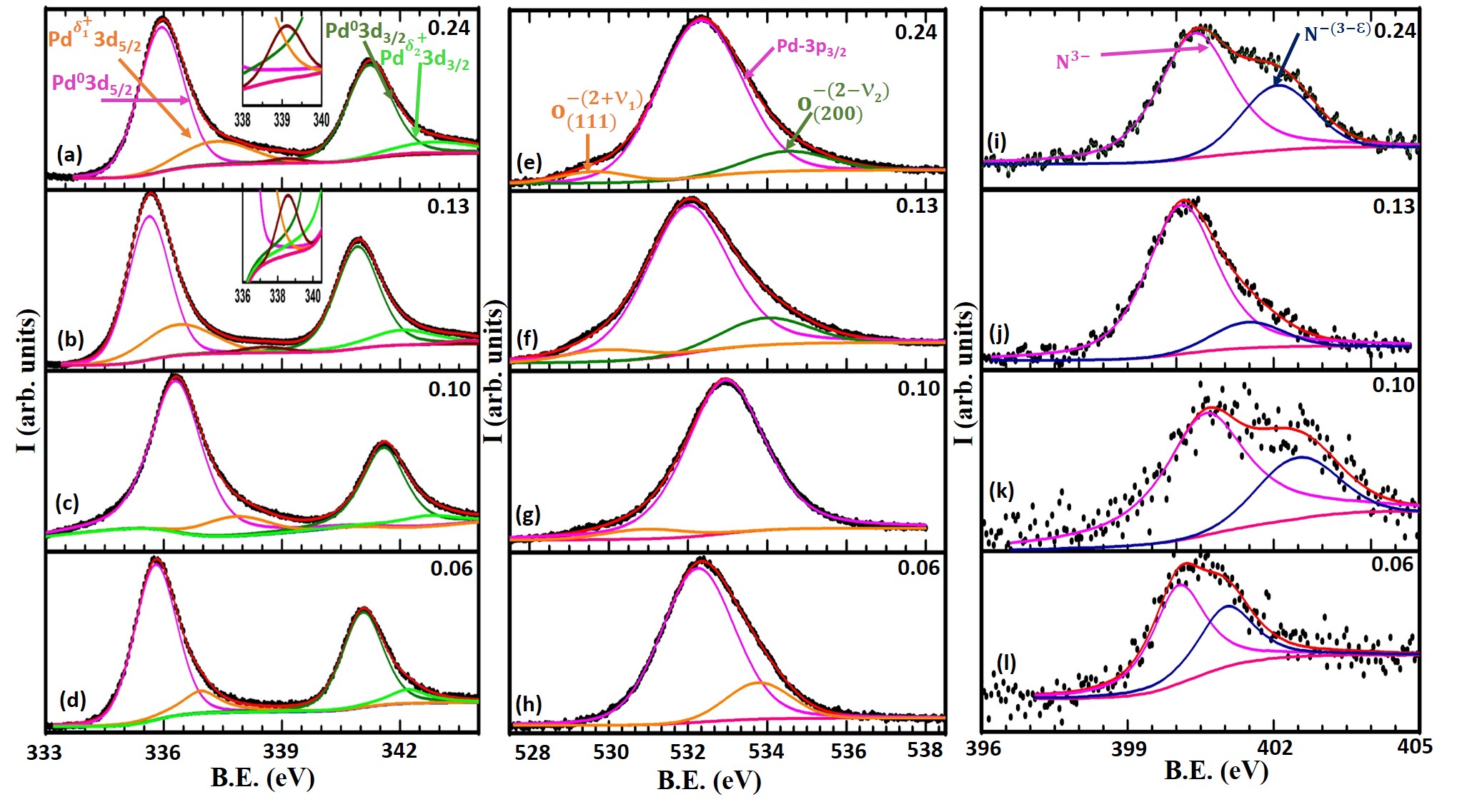}
	\caption{X-ray photoelectron spectroscopy measurements on the novel architecture with varying PVP$/$Pd mole ratios. (a)-(d) Pd-3d5$/$2 and Pd-3d3$/$2 core-level spectra for PVP$/$Pd mole ratio (a) 0.24, (b) 0.13, (c) 0.10 and (d) 0.06. Filled black circles in each figure (a)-(l) are the data points while the superimposed red solid curves on the data points corresponds to a fit using XPS Peak software. Resolved peaks in each figure are shown by thin coloured lines, where  Pd$^0$3d$_{5/2}$ (pink), Pd$^{\delta_1^+}$3d$_{5/2}$ (orange), Pd$^0$3d$_{3/2}$ (green) and Pd$^{\delta_1^+}$3d$_{3/2}$ (olive) in Figs. (a)-(d) are marked by arrows. Insets of (a) and (b) represent an expanded view of the data in the B.E. range of 336-340 eV to show the presence of another wine coloured resolved peak at (a) 339.1 eV and (b) 338.6 eV. (e)-(f) O-1s core-level spectra for PVP$/$Pd mole ratio (e) 0.24, (f) 0.13, (g) 0.10 and (h) 0.06. The resolved peaks corresponding to Pd-3p$_{3/2}$ (pink), O$^{-(2+\nu_1)}_{(111)}$ (orange) and O$^{-(2-\nu_2)}_{(200)}$ (olive) are marked by arrows. (i)-(l) N-1s core-level spectra for PVP$/$Pd mole ratio (i) 0.24, (j) 0.13, (k) 0.10 and (l) 0.06. The resolved peaks corresponding to N$^{3-}$ (pink) and N$^{\delta3-}$ (blue) are marked by arrows.}
	\label{fig:XPS}
\end{figure}

If surface adsorption of H atoms result in a resistive response of the PVP-Pd complex only for PVP$/$Pd ratio of 0.10, it must be happening that only for this ratio of PVP$/$Pd, the number of bare Pd sites on the PVP-Pd complex, $S_{Pd}$, is largest and sufficient for the phenomenon of surface adsorption of H atoms to take place. In order to confirm this, we did x-ray photoelectron spectroscopy (XPS) measurements, an extremely sensitive technique to observe core-level shifts in binding energies of elements according to their local environment, on all the architectures with varying ratios of PVP$/$Pd. Figs. \ref{fig:XPS} (a)-(d) plot the XPS spectra of Pd measured on devices with decreasing PVP$/$Pd ratio from 0.24 to 0.06 respectively. The spectra of Figs. \ref{fig:XPS} (a)-(d) is characterized by two peaks, typical of 3-d core levels governed by spin-orbit coupling. The first peak at a lower binding energy (B.E.) of $\sim$ 336 eV corresponds to the 3d-5$/$2 spin-orbit level of Pd while the second peak at a higher B.E. of $\sim$ 341.2 eV corresponds to that of 3d-3$/$2. The intensity ratios of the deconvoluted peaks (pink with respect to olive and orange with respect to light green) occur roughly in the expected ratio of 3:2 corresponding to the multiplicity of the spin-orbit split features. For brevity, we will only discuss the details of the 3d-5$/$2 spin-orbit level-the discussions on 3d-3$/$2 level follows exactly. For the PVP$/$Pd ratios of 0.24 and 0.13, the superimposed red curve was deconvoluted to reveal 3 peaks each for a given spin-orbit level. The main peak, shown in pink, occurs at 335.9 eV while the other two deconvoluted peaks, shown in orange and brown, occur at higher B.E.’s of 337.3 eV and 339.1 eV respectively. The main peak corresponds to palladium in the metallic state, Pd(0) \cite{xuxpspvp,garciacnt,garcia,collins}, in good agreement with bulk Pd (335.2 eV) value \cite{lundgren}. The slightly higher value of B.E. at 335.9 eV is known to be due to size and electronic effects \cite{paredes}. By studying the effect of chemisorption of oxygen atoms on the (111) surface of Pd nanocrystal, Ketteler et al. and Lundgren et al. \cite{ketteler,lundgren} found the B.E. to shift by + 1.4 eV from the main Pd(0) peak. So, the occurrence of the second deconvoluted peak (shown as orange) at a B.E. shift of exactly +1.4 eV is attributed to the chemisorption of oxygen on the (111) lattice plane of Pd nanocrystal that gets stabilized (refer to Fig. \ref{fig:TEM}(b)). From free energy minimization calculations, it was found that the (111) surface has the least energy followed by (100) surface followed by (110) surface \cite{mittendorfer,lundgren}. So, the second lower intensity peak at 339.1 eV is attributed to oxygen coordinated Pd (200) surface. The shift in B.E. at +3.2 eV is slightly higher than $\sim$ +2.2 eV values reported \cite{puecert,wang100}. However, it should be noted that the reported values were measured on Pd surface with only oxygen and not on a Pd-PVP composite structure, which may result in slight differences in binding energies. Since a higher B.E. is a measure of higher oxidation state of an atom, charge transfer between the chemisorbed PVP molecule and Pd crystal must happen in such a way that it leads to a withdrawal of electron density from Pd surface resulting in Pd$^{\delta^+}$ (Pd(II)). So, the two peaks at 337.3 eV and 339.1 eV have been nomenclatured as Pd$^{\delta_1^+}$ and Pd$^{\delta_2^+}$ respectively.\\
Presence of Pd$^{\delta_1^+}$ and Pd$^{\delta_2^+}$ spectral features in the XPS spectra of Pd indicates the poisoning of bare Pd metal sites since chemisorption of PVP ligand with Pd surface happens at these sites. So, the H$_2$ gas sensing ability of the PVP-Pd complex is critically dependent on the number of active Pd metal sites whose number can be estimated by the calculated area ratio of integrated intensity Pd$^{\delta_1^+}$. Higher the ratio, higher is the availability of the free active sites, S$_{Pd}$. Table \ref{table:arearatio} below describes the details of area ratios for Pd, O and N obtained from the XPS data.

\begin{table}[h]
	\centering
	\begin{tabular}{|c|c|c|c|c|}
		\hline
		S. No. & PVP/Pd ratio & Pd$^0/$Pd$^{\delta_1^+}$ & Pd-3p$_{3/2}/$O$^{-(2+\nu_1)}_{(111)}$& N$^{3-}/$N$^{-(3-\epsilon)}$ \\ \hline
		1 & 0.24 & 4.14:1 & 8.44:1 & 2.53:1  \\ \hline
		2 & 0.13 & 2.85:1 & 6.12:1 & 1.57:1 \\ \hline
		3 & 0.10 & 7.60:1 & 18.12:1 &  3.08:1 \\ \hline
		4 & 0.06 & 5.22:1 & 5.31:1  & 1.66:1 \\ \hline  
	\end{tabular}
	\caption{Pd$^0/$Pd$^{\delta 1+}$, Pd-3p$_{3/2}/$O$^{-(2+\nu_1)}_{(111)}$ and N$^{3-}/$N$^{-(3-\epsilon)}$ integrated area ratios obtained from XPS spectra for the novel architecture having varying PVP$/$Pd ratios.}
	\label{table:arearatio}
\end{table} 

%\begin{figure}%[tbhp]
%	\centering
%	\includegraphics[width=1\linewidth]{Fig6.jpg}
%	\caption{(Colour online) : O-1s core-level spectra for PVP/Pd mole ratio (a) 0.24, (b) 0.13, (c) 0.10 and (d) 0.06. Black filled circles correspond to the data points while the red envelope is a fit. Resolved peaks are shown by thin coloured lines.}
%	\label{fig:Oxps}
%\end{figure}

From the Table \ref{table:arearatio}, it can be seen that for the PVP$/$Pd ratio of 0.10, the integrated area ratio of Pd$^0$:Pd$^{\delta_1^+}$ is the maximum, suggesting that the number of available free Pd sites, S$_{Pd}$, is the maximum at PVP$/$Pd ratio of 0.10. So, the optimal PVP$/$Pd ratio for maximum available free Pd sites for adsorption of incoming H$_2$ gas molecules is achieved at 0.10. Additionally, for the PVP$/$Pd ratio of 0.24 and 0.13, the presence of additional Pd$^{\delta_2^+}$ spectral feature implies a further loss of bare active sites of Pd metal that may be available for incoming H$_2$ gas resulting in the absence of sensing in architectures of those ratios.\\
In order to understand the details of the PVP-Pd interaction, we studied the XPS spectra of oxygen and nitrogen shown in Figs. \ref{fig:XPS} (e)-(h) and Figs. \ref{fig:XPS} (i)-(l) respectively. Figs. \ref{fig:XPS} (e)-(h) show the O-1s spectra for PVP$/$Pd ratio varying from 0.24 to 0.06 respectively. The deconvoluted spectra reveals a main peak (shown as pink) at a B.E. of 532.2 eV for all the ratio’s barring that of 0.10, where it occurs at a B.E. of 533 eV. It is known that at this value of energy, the peak corresponding to Pd-(3p$_{3/2}$) overlaps with O-1s. However, since the amount of chemisorption of O on Pd is expected to be quite small, the very intense peak at 532.2 eV is ascribed to Pd-(3p$_{3/2}$). The deconvolution also shows the presence of additional peaks, two each for PVP$/$Pd ratios of 0.24 and 0.13 respectively, but only one peak for the PVP$/$Pd ratios of 0.10 and 0.06. It is known that when Pd forms an oxide on its (111) surface, it results in an occurrence of O 1s spectral feature at a B.E. at 529.8 eV \cite{lundgren}. So, the first additional peak at 529.8 eV is attributed to PdO formation on the (111) surface of Pd that gets stabilized (refer to Fig. \ref{fig:synthesisschematic}). Since the peak is at a lower binding energy as compared to O$^{2-}$, charge transfer happens from Pd to O to result in an excess charge on O, resulting in O$^{-(2+\nu_1)}_{(111)}$, where $\nu_1$ is positive and (111) denotes O binding on the (111) surface. The second spectral feature at 534.5 eV is attributed to PdO formation on the (200) surface of Pd \cite{puecert,wang100}, resulting in O$^{-(2-\nu_2)}_{(200)}$. So the O-1s spectra shows exactly similar features as the spectra of Pd shown in Figs. \ref{fig:XPS} (a)-(d) where two additional peaks were observed for the PVP$/$Pd ratio of 0.24 and 0.13 but only one additional peak for the spectra corresponding to PVP$/$Pd ratio of 0.10 and 0.06. However, the N-1s spectra shown in Figs. \ref{fig:XPS} (i)-(l) shows only one additional deconvoluted peak (shown in blue) between the B.E range of 401 and 402.5 eV, apart from the main N-1s peak (shown as pink) at a B.E. of $\sim$ 400.1 eV. The presence of an additional peak at a higher B.E. is a clear indication of chemisorption of N atom of the PVP ligand on the Pd surface, such that charge is transferred from N to Pd. The fraction of N$^{3-}/$N$^{-(3-\epsilon)}$, obtained from area ratio of integrated intensity N$^{3-}/$N$^{-(3-\epsilon)}$, is found to be maximum for PVP$/$Pd ratio of 0.10, in exact conformity with the results obtained from Pd’s spectra.
\begin{figure}[tbhp]
	\centering
	\includegraphics[width=0.8\linewidth]{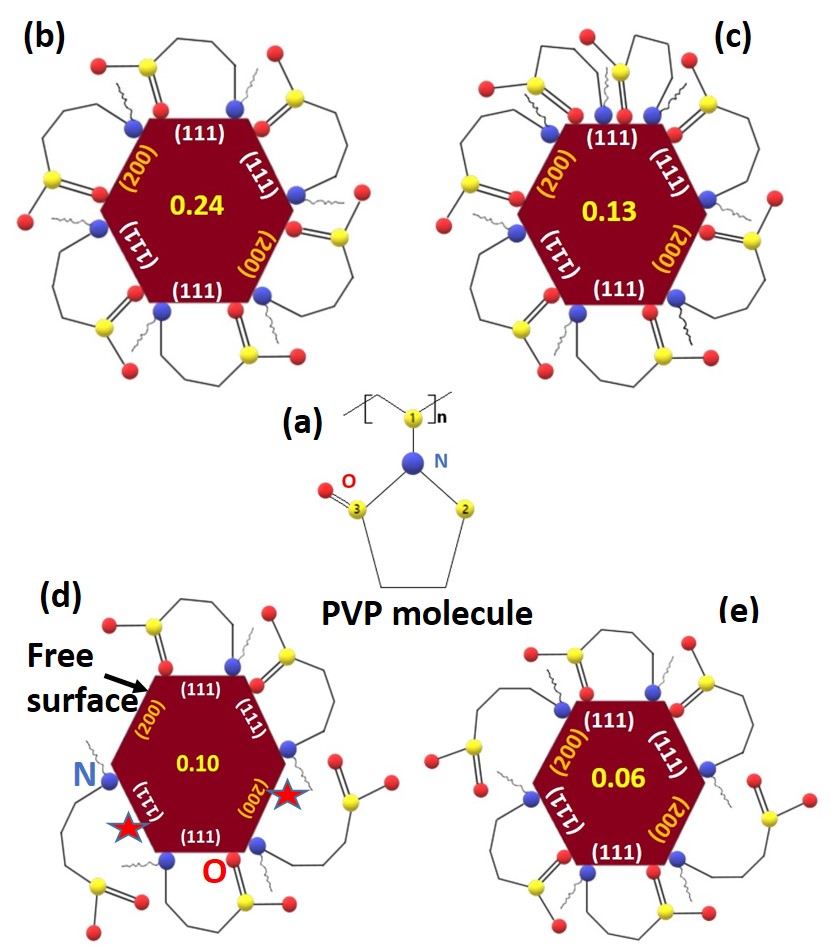}
	\caption{Schematic of Pd's chemisorption with PVP molecule in different architectures of varying PVP$/$Pd mole ratios. Schematic illustration of (a) PVP molecule, (b)-(e) Hexagonally shaped Pd nanocrystals with (111) and (200) surfaces having chemisorbed PVP molecules with their N (blue) atoms and O (red) atoms on various surfaces in architectures with PVP$/$Pd mole ratios of (b) 0.24, (c) 0.13, (d) 0.10 and (e) 0.06. Free surface is marked by an arrow while a free site is marked by a pink star in (d).}
	\label{fig:chemisorptionschematic}
\end{figure}

From the discussions of Fig. \ref{fig:XPS}, it is clear that PVP chemisorbs on the Pd nanocrystal on all the devices via its O and N atoms in such a way that oxygen atom chemisorbs on the Pd nanocrystal at both the (111) as well as the (200) surfaces on architectures that are prepared in the PVP$/$Pd mole ratios of 0.24 and 0.13, but it only chemisorbs on the (111) surface in the devices of PVP$/$Pd ratios of 0.10 and 0.06, resulting in an increase in the number of active sites for H$_2$ absorption for these ratios. Figs. \ref{fig:chemisorptionschematic} (b)-(e) provide such a schematic illustration where the chemisorption of the PVP molecules on the hexagonally shaped Pd nanocrystals are shown via PVP molecule's O and N atoms. It can be seen that while all the 6 surfaces (four (111) and two (200)) of Pd nanocrystals are occupied by PVP molecules for the architectures with PVP$/$Pd mole ratios of 0.24 and 0.13 (see Fig. \ref{fig:chemisorptionschematic} (b) and (c) respectively), those with mole ratios of 0.10 and 0.06 have free surfaces and sites. For instance, O atoms are not shown to be bound to Pd (200) surfaces of architectures represented in Figs. \ref{fig:chemisorptionschematic} (d) and (e).  However, since Pd, O and N XPS' area ratios are higher for architecture of 0.10 compared to 0.06, the number of free active sites, S$_{Pd}$ is higher for PVP$/$Pd ratio of 0.10. Since the H$_2$ loading happens via the (111) surface which is a very fast process \cite{johnson}, we observe extremely good rise time constants and sensitivity with the architecture of PVP$/$Pd mole ratio of 0.10. However, since the H$_2$ unloading happens via the (200) surface which is a slow process \cite{johnson}, the recovery times (see Figs. \ref{fig:hydrogensensing}) (a)-(b) is not as fast.

\begin{figure}[tbhp]
	\centering
	\includegraphics[width=1.0\linewidth]{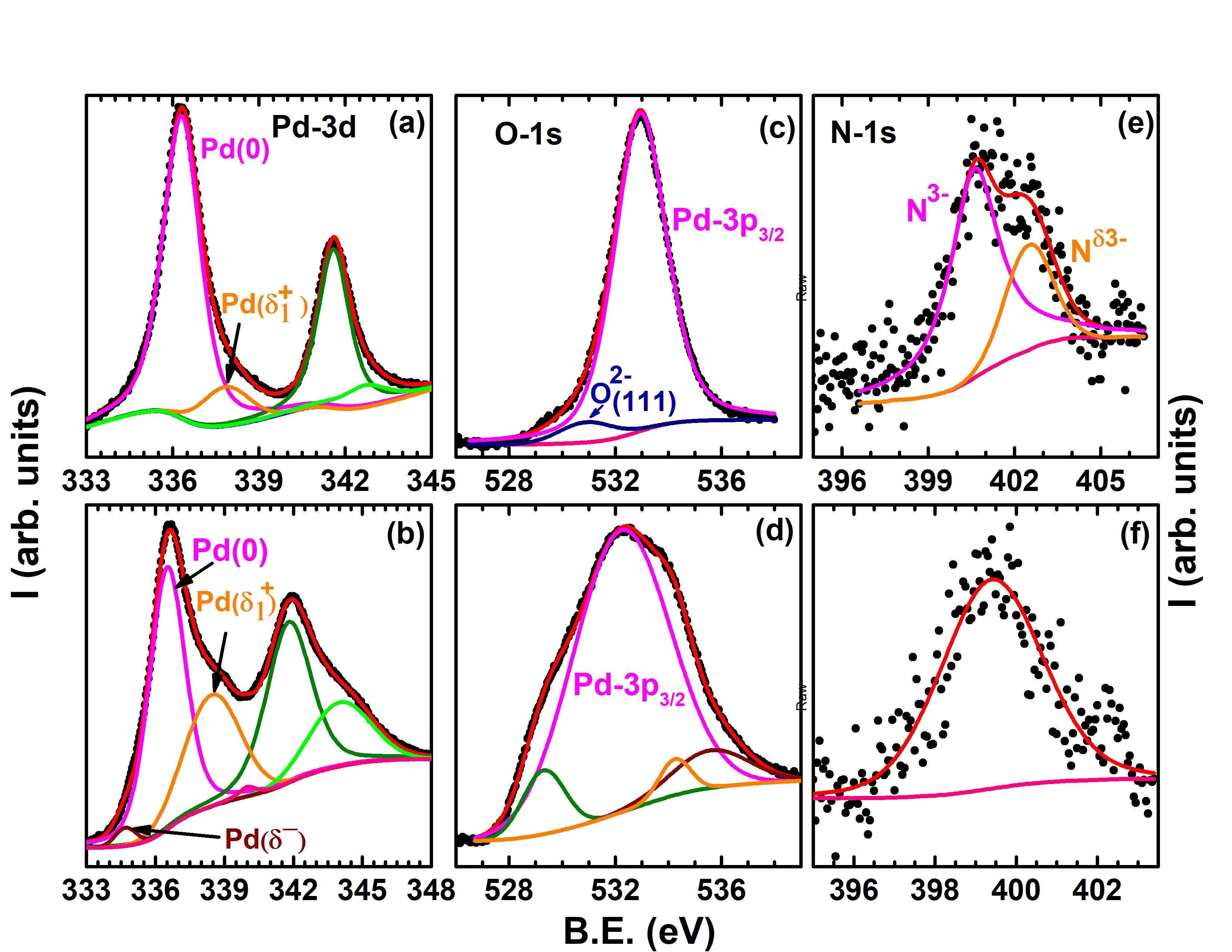}
	\caption{Change of XPS spectra of Pd, O, N on the architecture with PVP$/$Pd mole ratio of 0.10 after ozone cleaning. (a), (c) and (e) represent XPS spectra of Pd, O and N atoms respectively before ozone cleaning while (b), (d) and (f) represent the corresponding spectra after ozone cleaning in an architecture having PVP$/$Pd mole ratio of 0.10. Occurence of new peaks as well as intensity increase of peaks compared to no ozone cleaning is clearly visible in Pd and O spectra.}
	\label{fig:ozonecleaning}
\end{figure}

In an effort to further increase the number of bare active sites S$_{Pd}$ on the Pd surface in the architecture with maximum sites, namely, with PVP$/$Pd mole ratio of 0.10, the architecture was subjected to UV-Ozone cleaning \cite{crespo}, a technique frequently used to remove compounds on the surface of nanoparticles without changing the morphological characteristics of the nanoparticles. However, it was found that the UV-cleaned architecture loses its H$_2$ gas sensing capability completely! To understand this rather strange result, we measured XPS on the UV-cleaned architecture. Fig. \ref{fig:ozonecleaning} (b) shows the Pd spectra after it was subjected to UV cleaning. The change in the spectra without any UV cleaning (shown in \ref{fig:ozonecleaning} (a)) is remarkable! It can be seen that not only has the intensity of the Pd$^{\delta_1^+}$ peak increased substantially compared to the uncleaned one, the spectra also shows the presence of another peak at a lower B.E. of 334.7 eV, suggesting the presence of a Pd$^{\delta -}$ species in the UV-cleaned sample. Additionally, the ratio of Pd(0)/Pd$^{\delta_1^+}$ was found to reduce substantially to 1.27 from the original 7.6 (see Table \ref{table:arearatio}) indicating the loss of bare Pd metal sites, which is further confirmed by the presence of the new Pd$^{\delta -}$ peak. However, the change in the O-1s spectra was found to be the most spectacular, as evident from Fig. \ref{fig:ozonecleaning} (d) where the O-1s spectra is now seen to have two additional peaks compared to the spectra with no UV cleaning (\ref{fig:ozonecleaning} (c)). So, the effect of oxygen cleaning is to poison the bare Pd sites via O bonding on the Pd sites, resulting in a loss of free active sites for incoming H$_2$ molecules. Additionally, the earlier N bonded sites get replaced by O atoms as evident from the vanishing additional peak of the initial N-1s spectra (Fig. \ref{fig:ozonecleaning} (e)) in the UV cleaned N-1s spectra (Fig. \ref{fig:ozonecleaning} (f)). 

\section*{Conclusions and Outlook}  
To conclude, we have been able to make a novel chemiresistive hybrid hydrogen sensor based on palladium nanocrystals and poly (N-vinyl-2-pyrrolidone) (PVP) polymer that was grown on a siloxane self-assembled monolayer to reduce the stiction of hydrogen on glass substrate. The novel architecture was found to have very good hydrogen sensing properties, with a high sensitivity of $\sim$ 30$\%$ for a 0.5$\%$ H$_2$ gas and a fast rise response time of $\sim$ 50 s. The value of obtained sensitivity is similar to the ones obtained using physical deposition method, making our chemiresistive sensor extremely cost-effective. A log-log plot of time-constant and concentration revealed the diffusion of H atoms inside the Pd lattice to be a very fast process. Detailed x-ray photoelectron spectroscopy measurements revealed the existence of largest number of bare palladium metal available sites and surfaces only for a given PVP$/$Palladium mole ratio of 0.10. An effort to increase the number of available sites by UV-ozone cleaning led to a degradation of the sensing device due to poisoning of the already available sites by O.\\
It is clear that the exceptional sensitivity and rise time-constants that have been achieved in our novel hybrid chemiresistive sensor holds great potential for future sensor technology since the given response occur at room temperature and atmospheric pressure conditions, the ambient conditions in real life applications of a low concentration H$_2$ leak detector or medical applications of breadth H$_2$ sensor. The exceptional sensing characteristics were found to arise due to a combination of various factors: (i) small size of the synthesised palladium nanocrystals with well-defined facets, (ii) fast diffusion of H atoms from the surface of the palladium nanocrystals onto the bulk and (iii) H$_2$ loading on the (111) surface that is known to be a very fast process. In a wider perspective, our results hold promise for next generation thin film based chemiresistive sensors where variation of the type of self-assembled monolayer as well as palladium to polymer ratios may vary not only the size but also the morphology of the grown nanocrystals, wherein, more facets of (111) could be grown. Such tailoring of properties may lead to thin-film based sensors that are safe, compact, easy to fabricate, are cost-effective and have enhanced sensitivity and low response times to meet the ever increasing demands of low concentration hydrogen gas sensors at ambient conditions of temperatures and pressures. 
%\authordeclaration{}

\section*{Acknowledgements}
P.S.G would like to thank IISER TVM for Institute post-doctoral fellowship and ISRO RESPOND (Grant No. ISRO/RES/3/762/17-18) for financial support. D.J-N acknowledges financial support from ISRO RESPOND, Govt. of India (Grants Nos. ISRO/RES/3/704/16-17 and ISRO/RES/3/762/17-18)

\section*{Author Contributions}
D.J-N. and P.S.G. designed research; C.E.A., P.S.G., A.J., A.K., N.S., N.K., V.B.K., S.M.N and D.J-N performed research; D.J-N analysed the data and wrote the manuscript with inputs from all co-authors.

%\section*{Competing Interests}
%The authors declare no competing interests.

\section*{Additional Information}
An Indian patent titled ``A chemiresistive hydrogen sensor and a method thereof'' with number 202041035199 has been filed based on this work.

%\subsection*{References}
%\bibliography{49}

\begin{thebibliography}{99}

\bibitem{yurum} Y{\"u}r{\"u}m, Y. Hydrogen energy system: production and utilization of hydrogen and future aspects. (Kluwer Academic Publishers, 295 Boston, 1995). 

%\bibitem{weast} Weast, R. C. in \textit{Handbook of Chemistry and Physics, Ed. 56} (CRC Press, Inc., Boca Raton, 1976).

%\bibitem{lundstrom} Lundstrom, I., Armgarth, M. \& Petersson, L.-G. Physics with catalytic metal gate chemical sensors. \textit{CRC Crit. Rev. Solid State Mater. Sci.} \textbf{15}, 201-278 (1989).

\bibitem{christofides} Christofides, C.; Mandelis, A. J. Solid‐state sensors for trace hydrogen gas detection. \textit{Appl. Phys.} \textbf{1990}, 68, R1-R30.

\bibitem{hubert} Hubert, T.; Boon-Brett, L.; Black, G.; Banach, U. Hydrogen Sensors a Review. \textit{Sens. Actuators B} \textbf{201}1, 157, 329-352.

\bibitem{rizkalla} Rizkalla, S. W.; Luo, J.; Kabir, M.; Chevalier, A.; Pacher, N.; Slama, G. Chronic consumption of fresh but not heated yogurt improves breath-hydrogen status and short-chain fatty acid profiles: a controlled study in healthy men with or without lactose maldigestion. \textit{Am. J. Clin. Nutr.} \textbf{2000}, 72, 1474-1479. 

%\bibitem{riordan} Riordan, S. M., McIver, C. J., Duncombe, V. M., Thomas, M. C. \& Bolin, T. D. Evaluation of the rice breath hydrogen test for small intestinal bacterial overgrowth. \textit{Am. J. Gastroenterol.} \textbf{95(10)}, 2858-2864 (2000).

%\bibitem{backus} Backus, R. C., Puryear, L. M., Crouse, B. A., Biourge, V. C., \& Rogers, Q. R. Breath hydrogen concentrations of cats given commercial canned and extruded diets indicate gastrointestinal microbial activity vary with diet type. \textit{J. Nutr.} \textbf{6}, 1763S-1766S Suppl. 2 (2002).

%\bibitem{burge} Burge, M. R., Tuttle, M. S., Violett, J. L., Stephenson, C. L. \& Schade, D. S. Breath hydrogen testing identifies patients with diabetic gastroparesis. \textit{Diabetes Care} \textbf{23(6)}, 860-861 (2000).

\bibitem{hunter} Hunter, G. W.;  Chen, L. Y.; Neudeck, P. G.; Knight, D.; Liu, C. C.; Wu,  Q. H.; Zhou, H. J.; Makel, D.; Liu, M.; Rauch, W. A. Chemical Gas Sensors for Aeronautic and Space Applications II, NASA/TM-208504 \textbf{1998}.

%\bibitem{hunter1} Hunter, G. W., Neudeck, P. G., Jefferson, G. D., Madzsar, G. C., Liu C. C. \& Wu Q. H. The Development of Hydrogen Sensor Technology at NASA Lewis Research Center, NASA/TM- 106141 (1992).

\bibitem{krall} Krall, J.; Glocer, A.; Fok, M.-C.; Nossal, S. M.; Huba, J. D. The unknown hydrogen exosphere: Space weather implications. \textit{Space Weather} 2018, 16, 205–215.

%\bibitem{yang} Yang, D., Valent{\'i}n, L., Carpena, J., Ota$\tilde{n}$o, W., Resto O. \& Fonseca, L. F. Temperature-Activated Reverse Sensing Behavior of Pd Nanowire Hydrogen Sensors. \textit{Small} \textbf{9}, 188 (2013).  

%\bibitem{jiang} Jiang, H., Yu, Y., Zhang, L., Zhu, J., Zhao, X. \& Zhang, W. Flexible and Highly Sensitive Hydrogen Sensor Based on Organic Nanofibers Decorated by Pd Nanoparticles. \textit{Sensors} \textbf{19}, 1290 (2019).

\bibitem{wang} Wang, Z.; Li, Z.; Jiang, T.; Xu, X.; Wang, C. Ultrasensitive hydrogen sensor based on Pd(0)-Loaded SnO$_2$ electrospun nanofibers at room temperature. \textit{ACS Appl. Mater. Interfaces} \textbf{2013}, 5, 2013-2021.

\bibitem{zeng} Zeng, X. Q.; Latimer, M. L.; Xiao, Z. L.; Panuganti, S.; Welp, U.; Kwok, W. K.; Xu, T. Hydrogen Gas Sensing with Networks of Ultrasmall Palladium Nanowires Formed on Filtration Membranes. \textit{Nano Lett.} \textbf{2011}, 11, 262-268.

\bibitem{jeon} Jeon, K. J.; Jeun, M.; Lee, E.; Lee, J. M.; Lee, K.-I.; Allmen, P.v.; Lee, W. Finite size effect on hydrogen gas sensing performance in single Pd nanowires. \textit{Nanotechnology} \textbf{2008}, 19, 495501-6. 

%\bibitem{walter} Walter, E. C., Favier, F. \& Penner, R. M. Palladium Mesowire Arrays for Fast Hydrogen Sensors and Hydrogen-Actuated Switches. \textit{Anal. Chem.} \textbf{74} 1546 (2002). 

%\bibitem{yun} Yun, M., Myung, N. V., Vasquez, R. P., Lee, C., Menke, E. \&  Penner, R. M. Electrochemically Grown Wires for Individually Addressable Sensor Arrays. \textit{Nano Lett.} \textbf{4} 419 (2004).

\bibitem{shin} Shin, D. H.; Lee, J. S.; Jun, J.; An, J. H.; Kim, S. G.; Cho, K. H.; Jang, J. Flower-like Palladium Nanoclusters Decorated Graphene Electrodes for Ultrasensitive and Flexible Hydrogen Gas Sensing. \textit{Sci. Rep.} \textbf{2015}, 5, 12294-11 (2015). 

%\bibitem{bucur} Bucur, R. V., Mecca, V. \& Flanagan, T. B. The kinetics of hydrogen (deuterium) sorption by thin palladium layers studied with a piezoelectric quartz crystal microbalance \textit{Surf. Sci.} \textbf{54}, 477-488 (1976).

%\bibitem{shaver} Shaver, P. J. Bimetal Strip Hydrogen Gas Detectors. \textit{Rev. Sci. Instrum.} \textbf{40}, 901-905 (1969).

%\bibitem{bevenot} B{\'e}venot, X., Trouillet, A., Veillas, C., Gagnaire, H., \& Clement, M. Hydrogen leak detection using an optical fibre sensor for aerospace applications. \textit{Sens. Actuators B} \textbf{67}, 57-67 (2000).

%\bibitem{dwivedi} Dwivedi, D., Dwivedi, R. \& Srivastava, S. K. Sensing properties of palladium-gate MOS (Pd-MOS) hydrogen sensor-based on plasma grown silicon dioxide. \textit{Sens. Actuators B} \textbf{71}, 161-168 (2000).

\bibitem{xu} Xu, T.; Zach, M. P.; Xiao, Z. L.; Rosenmann, D.; Welp, U.; Kwok, W. K.; Crabtree, G. W. Self-assembled monolayer-enhanced hydrogen sensing with ultrathin palladium films. \textit{Appl. Phys. Lett.} \textbf{2005}, 86, 203104-3.

\bibitem{huang} B-R Huang; J-C Lin. Core–shell structure of zinc oxide/indium oxide nanorod based hydrogen sensors. \textit{Sens. Actuators B} \textbf{2012}, 174, 389-393.

%\bibitem{dankert} Dankert, O. \& Pundt, A. Hydrogen-induced percolation in discontinuous films. \textit{Appl. Phys. Lett.} \textbf{81}, 1618-1620 (2002).

\bibitem{lee} Lee, J-S; Seo, M-H; Choi, K-W; Yoo, J-Y; Joa M-S; Yoon, J-B.  Stress-engineered palladium nanowires for wide range (0.1\%–3.9\%) of H$_2$ detection with high durability. \textit{Nanoscale} \textbf{2019}, 11, 16317-16326 (2019).

\bibitem{lee1} Lee, E.; Lee, J. M.; Koo, J. H.; Lee, W.; Lee, T. Hysteresis behavior of electrical resistance in Pd thin films during the process of absorption and desorption of hydrogen gas. \textit{I. Journal. Hydrogen Energy} \textbf{2010}, 35, 6984-6991.

\bibitem{kumar} Kumar, M. K.; Rao, M. S. R.; Ramaprabhu, S. Structural, morphological and hydrogen sensing studies on pulsed laser deposited nanostructured palladium thin films. \textit{J. Phys. D: Appl. Phys.} \textbf{2006}, 39, 2791-2975.

\bibitem{ramanathan} Ramanathan, M.; Skudlarek, G.; Wang H. H.; Darling, S. B. Crossover behavior in the hydrogen sensing mechanism for palladium ultrathin films. \textit{Nanotechnology} \textbf{2010}, 21, 125501-6.

\bibitem{ibanez} Iba$\tilde{n}$ez F. J.; Zamborini, F. P. Ozone- and Thermally Activated Films of Palladium Monolayer-Protected Clusters for Chemiresistive Hydrogen Sensing. \textit{Langmuir} 2006, 22, 9789-9796.

\bibitem{noh} Noh, J-S; Lee J. M.; Lee, W. Low-Dimensional Palladium Nanostructures for Fast and Reliable Hydrogen Gas Detection. \textit{Sensors} \textbf{2011}, 11, 825-851.

%\bibitem{sakamoto} Sakamoto, Y., Takai, K., Takashima I. \% Imada, M. Electrical resistance measurements as a function of composition of palladium–hydrogen(deuterium) systems by a gas phase method. \textit{J. Phys.: Condens. Matter} \textbf{8} 3399-3411 (1996).

\bibitem{kay} Kay, B. D.; Peden, C. H. F.; Goodman, D. W. Kinetics of hydrogen absorption by Pd(110). \textit{Phys. Rev. B} \textbf{1986}, 34, 817-822.

\bibitem{lewis} Lewis, F. A. The Palladium-Hydrogen system: Structures near phase transition and critical points. \textit{Int. J. Hydrogen Energy} \textbf{1995}, 20, 587-592.

%\bibitem{wolf} Wolf, J., Lee, M. W., Davis, R. C., Fay, P. J. \& Ray, J. R. Pressure-composition isotherms for palladium hydride. \textit{Phys. Rev. B} \textbf{48}, 12415-12418 (1993).

\bibitem{flanagan} Flanagan B.; Oates, W. A. The Palladium-Hydrogen system. \textit{Annu. Rev. Mater. Sci.} \textbf{1991}, 21, 269-304.

\bibitem{dharmendra} Singh, D. K.; Praveen, S. G.; Kamble, V. B.; Mitra, J.; Jaiswal-Nagar, D. Thickness induced metal to insulator charge transport and unusual hydrogen response in granular palladium nanofilms (submitted).

\bibitem{favier} Favier, F.; Walter, E. C.; Zach, M. P.; Benter, T.; Penner, R. M. Hydrogen sensors and switches from electrodeposited Palladium mesowire arrays. \textit{Science} \textbf{2001}, 293, 2227-2231.

%\bibitem{kaltenpoth} Kaltenpoth, G., Schnabel, P., Menke, E., Walter, E. C.,  Grunze, M. \& Penner, R. M. Multimode Detection of Hydrogen Gas Using Palladium-Covered Silicon $\mu$Channels. \textit{Anal. Chem.} \textbf{75}, 4756-4765 (2003).

\bibitem{teranishi} Teranishi T.; Miyake, M. Size Control of Palladium Nanoparticles and their Crystal Structures. \textit{Chem. Mater.} \textbf{1998}, 10, 594-600.

%\bibitem{cookson} Cookson, J. The Preparation of Palladium Nanoparticles. \textit{Platinum Metals Rev.} \textbf{56(2)}, 83–98 (2012).

\bibitem{collins} Collins, G.; Schmidt, M.; McGlacken, G. P.; O{'}Dwyer, C; Holmes, J. D. Stability, Oxidation, and Shape Evolution of PVP-Capped Pd Nanocrystals. \textit{J. Phys. Chem. C} \textbf{2014}, 118, 6522-6530. 

\bibitem{garcia} Garc{\'i}a-Aguilar, J.; Navlani-Garc{\'i}a, M.; Berenguer-Murcia, {\'A}.; Mori, K.; Kuwahara, Y.; Yamashita, H.; Cazorla-Amoros, D. Evolution of the PVP-Pd surface interaction in nanoparticles through the case study of formic acid decomposition. \textit{Langmuir} \textbf{2016}, 32, 12110-12118.

\bibitem{grammatikopoulos} Grammatikopoulos, P.; Cassidy, C.; Singh V.; Sowwan, M. Coalescence-induced crystallization wave in Pd nanoparticles. \textit{Sci. Rep.} \textbf{2014}, 4, 5779:1-9.

%\bibitem{hirai} Hirai, H. \& Toshima, N. in \textit{Tailored Metal Catalysts} (Ed.  Iwasawa, Y.) 87 (Reidel, Dordrecht, 1986).

\bibitem{xian} Xian, J.; Hua, Q.; Jiang, Z.; Ma, Y.; Huang, W. Size-Dependent Interaction of the Poly(N-vinyl-2-pyrrolidone) Capping Ligand with Pd Nanocrystals, \textit{Langmuir} \textbf{2012}, 28, 6736-6741.

\bibitem{suresh} Suresh, S.; Kusuma, U.; Anupama, T. V.; Sriram, S.; Kamble, V. Analysis of unusual and instantaneous overshoots in response transients of gas sensors. \textit{Phys. Status Solidi-RRL} \textbf{2019}, 3, 1800683-7.

\bibitem{teranishi1} Teranishi, T.; Nakata, K.; Iwamoto, M.; Miyake, M.;  Toshima, N. Promotion effect of lanthanoid ions on catalytic activity of polymer-immobilized palladium nanoparticles. \textit{React. Funct. Polym.} \textbf{1998}, 37, 111-119.

\bibitem{sil} Sil, D.; Hines, J.; Udeoyo, U.; Borguet, E. Palladium Nanoparticle-Based Surface Acoustic Wave Hydrogen Sensor. \textit{ACS Appl. Mater. Inter.} \textbf{2015}, 7, 5709-5714. 

\bibitem{berger} Berger, D.; Trǎistaru, G. A.; Vasile, B. Ş.; Jitaru, I.; Matei, C. Palladium nanoparticles synthesis with controlled morphology obtained by polyol method. \textit{U.P.B. Sci. Bull. Series B} \textbf{2010}, 72, 113-120.

\bibitem{kockzur} Koczkur, K. M.; Mourdikoudis, S.; Polavarapu, L.; Skrabalak, S. E. Polyvinylpyrrolidone (PVP) in nanoparticle synthesis. \textit{Dalton Trans.} \textbf{2015}, 44, 17883-17905.

\bibitem{gardner} Gardner J. W. A Non-linear Diffusion reaction Model of Electrical Conduction in Semiconductor Gas Sensors. \textit{Sens. Actuators B Chem.} \textbf{1990}, 1, 166-170.

\bibitem{yadav} Yadav, S.; Nair, A.; Kusuma, U. M. B.; Kamble, V. B. Protonic Titanate Nanotube–Reduced Graphene Oxide Composites for Hydrogen Sensing \textit{ACS Appl. Nano Mater.} DOI: 10.1021/acsanm.0c02077

\bibitem{johnson} Johnson, N. J. J.; Lam, B.; MacLeod, B. P.; Sherbo, R. S.;
Moreno-Gonzalez, M.; Fork, D. K.; Berlinguette, C. P. Facets and vertices regulate hydrogen uptake and release in palladium nanocrystals. \textit{Nat. Mat.} \textbf{2019}, 18, 454-458.

\bibitem{zalineeva} Zalineeva, A.; Baranton, S.; Coutanceau, C.; Jerkiewicz, G. Octahedral palladium nanoparticles as excellent hosts
for electrochemically adsorbed and absorbed hydrogen. \textit{Sci. Adv.} \textbf{2017}, 3, 1600542-10.
%\bibitem{pooja} Pooja, Barman, P. B., Hazra, S. K. Role of Capping Agent in Palladium Nanoparticle Based Hydrogen Sensor. \textit{J. Clust. Sci.} \textbf{29}, 1209-1216 (2018).

\bibitem{sun} Sun, Y.; Wang, H. High-Performance, Flexible Hydrogen Sensors That Use Carbon Nanotubes Decorated with Palladium Nanoparticles. \textit{Adv. Mater.} \textbf{2007}, 9, 2818–2823.

\bibitem{sakaguchi} Sakaguchi, H.; Yagi, Y.; Taniguchi, N.; Adachi, G.; Shiokawa, J. Effects of Hydrogen absorption on the electrical resistivity of LaCo, films and the determination of the hydrogen content in the films. \textit{J. Less Common Met.} \textbf{1987}, 135, 137-146. 

\bibitem{sieverts} Sieverts, A. Absorption of Gases by Metals. \textit{Z. Metallkund.} \textbf{1929}, 21, 37-46.

\bibitem{mueller} Mueller, W. M.; Blackledge, J. P.; Libowitz, G. G. \textit{Metal Hydrides} (Academic Press: New York, 1968).
	
\bibitem{barr} Barr, A. The effect of Hydrogen absorption on the electrical conduction in discontinuous Palladium films. \textit{Thin Solid Films} \textbf{1977}, 41, 217-226.	

\bibitem{wu} Wu F.; Morris, J. E. The effects of hydrogen absorption on the electrical conduction in discontinuous palladium films. \textit{Thin Solid Films} \textbf{1994}, 246, 17-23.

\bibitem{morris} Morris, J. E.; Kiesow, A.; Hong, M.; Wu, F. Effects of hydrogen absorption on the electrical conduction of discontinuous palladium thin films. \textit{Int. J. Electronics} \textbf{1996}, 81, 441-447.

%\bibitem{xpspeak} http//www.phy.cuhk.edu.hk/~surface/XPSPEAK/

\bibitem{xuxpspvp} Xu, H.; Ding, L.-X.; Liang, C.-L.; Tong, Y.-X.; Li, G.-R.  High-performance polypyrrole functionalized PtPd electrocatalysts based on PtPd/PPy/PtPd three-layered nanotube arrays for the electrooxidation of small organic molecules. \textit{NPG Asia Materials} \textbf{2013}, 5, 69:1-10.

\bibitem{garciacnt} Garc{\'i}a-Aguilar, J.; Miguel-Garc{\'i}a, I.; Murcia, A. B.; Cazorla-Amor{\'o}s, D. Single wall carbon nanotubes loaded with Pd and NiPd nanoparticles for H$_2$ sensing at room temperature. \textit{Carbon} \textbf{2014}, 66, 599–611. 

\bibitem{lundgren} Lundgren, E.; Kresse, G.; Klein, C.; Borg, M.; Andersen, J. N.; De Santis, M.; Gauthier, Y.; Konvicka, C.; Schmid, M.; Varga, P. Two-Dimensional Oxide on Pd (111). \textit{Phys. Rev. Lett.} \textbf{2002}, 88, 246103-246104.

%\bibitem{smirnov} Smirnov, M. Y., Vovk, E. I., Kalinkin, A. V., Pashis, A. V. \& Bukhtiyarov, V. I. An XPS Study of the Oxidation of Noble Metal Particles Evaporated onto the Surface of an Oxide Support in their Reaction with NO$_x$. \textit{Kinet. Catal.} \textbf{53}, 117-124 (2012). 

\bibitem{paredes} Paredis, K.; Ono, L. K.; Behafarid, F.; Zhang, Z.; Yang, J. C.; Frenkel, A. I.; Cuenya, B. R. Evolution of the Structure and Chemical State of Pd Nanoparticles during the in Situ Catalytic Reduction of NO with H$_2$. \textit{J. Am. Chem. Soc.} \textbf{2011}, 133, 13455-13464.

\bibitem{ketteler} Ketteler, G.; Ogletree, D. F.; Bluhm, H.; Liu, H. J.; Hebenstreit, E. L. D.; Salmeron, M. In Situ Spectroscopic Study of the Oxidation and Reduction of Pd(111). \textit{J. Am. Chem. Soc.} \textbf{2005}, 127, 18269-18273.

\bibitem{mittendorfer} Mittendorfer, F.; Seriani, N.; Dubay, O.; Kresse, G.  Morphology of mesoscopic Rh and Pd nanoparticles under oxidizing conditions. \textit{Phys. Rev. B} \textbf{2007}, 76, 233413-4.

\bibitem{puecert} Puecert, M. XPS Study on Surface and Bulk Palladium Oxide, Its Thermal Stability, and a Comparison with Other Noble Metal Oxides. \textit{J. Phys. Chem.} \textbf{1985}, 89, 2481-2486.

\bibitem{wang100} Wang, J.; Yun, Y.; Altman, E. I. The plasma oxidation of Pd (100). \textit{Surf. Sci.} \textbf{2007}, 601, 3497–3505.

\bibitem{crespo} Crespo-Quesada, M.; Andanson, J.-M.; Yarulin, A.; Lim, B.; Xia, Y.; Kiwi-Minsker, L. UV-Ozone Cleaning of Supported Poly(vinylpyrrolidone)-Stabilized Palladium Nanocubes: Effect of Stabilizer Removal on Morphology and Catalytic Behavior. \textit{Langmuir} \textbf{2011}, 27, 7909–7916.

\end{thebibliography}

\end{document}

\begin{abstract}
	In this supplementary, we provide details of (I) TEM images for all the architectures with varying PVP$/$Pd mole ratio, (II) R-t sensing graphs to show repeatability and (III) Comparison of sensing performance of various Pd based resisitve sensing devices. 
\end{abstract}
\date{\today}
\maketitle

\section{TEM imaging for all architectures}
\begin{figure}
	\centering
	\includegraphics[width=1\linewidth]{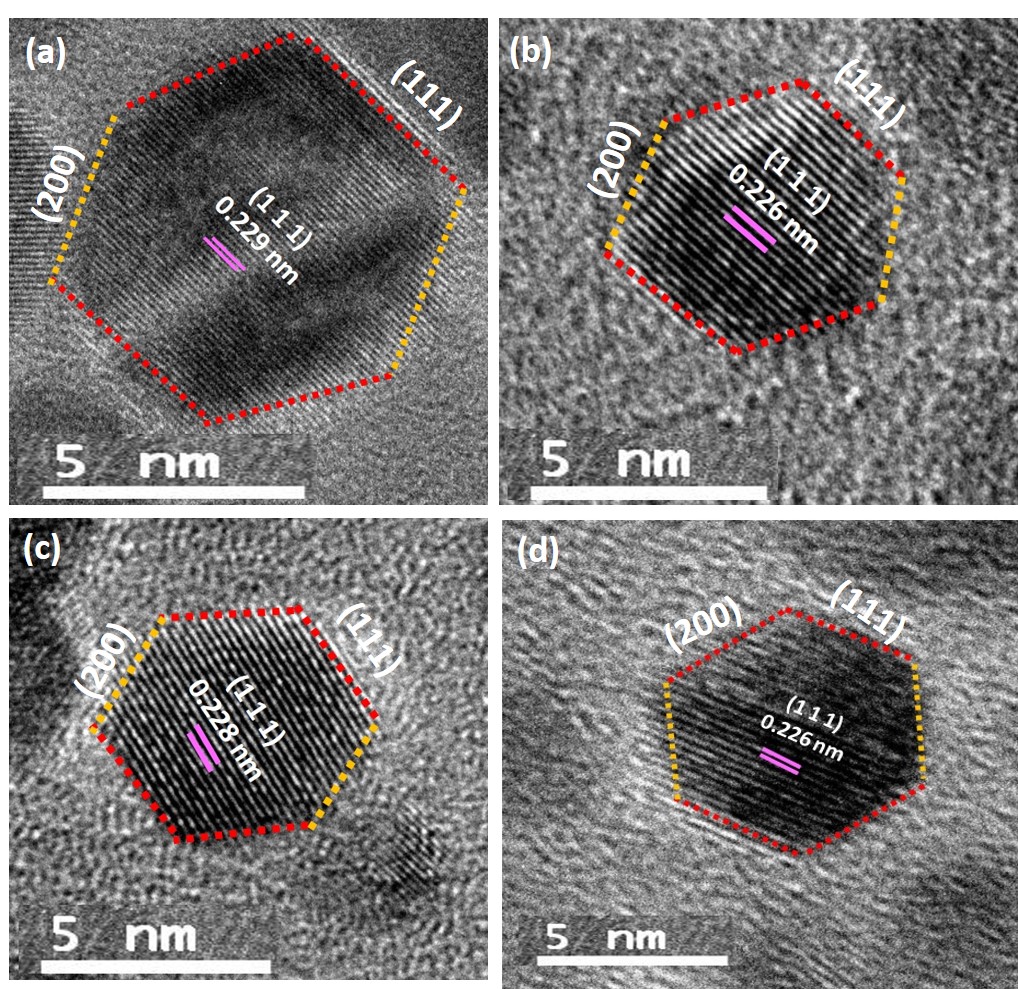}
	\caption{(Colour online): TEM images for the architectures with PVP$/$Pd mole ratio (a) 0.06, (b) 0.10, (c) 0.13 and (d) 0.24.}
	\label{fig:TEM}
\end{figure}

In the main text, we have shown that the structure of the nanocrystal is hexagonal in shape for the architecture of PVP$/$Pd mole ratio of 0.10. This shape is exhibited by all the architectures with PVP$/$Pd mole ratios ranging from 0.06-0.24 as shown in Figs. \ref{fig:TEM} (a)-(d) respectively. Interplanar separation corresponding to the (111) plane \cite{grammatikopoulos} has been marked clearly for all the planes. (200) planes have also been marked for all the architectures.

\section{Sensing measurements}

\begin{figure}
	\centering
	\includegraphics[width=1\linewidth]{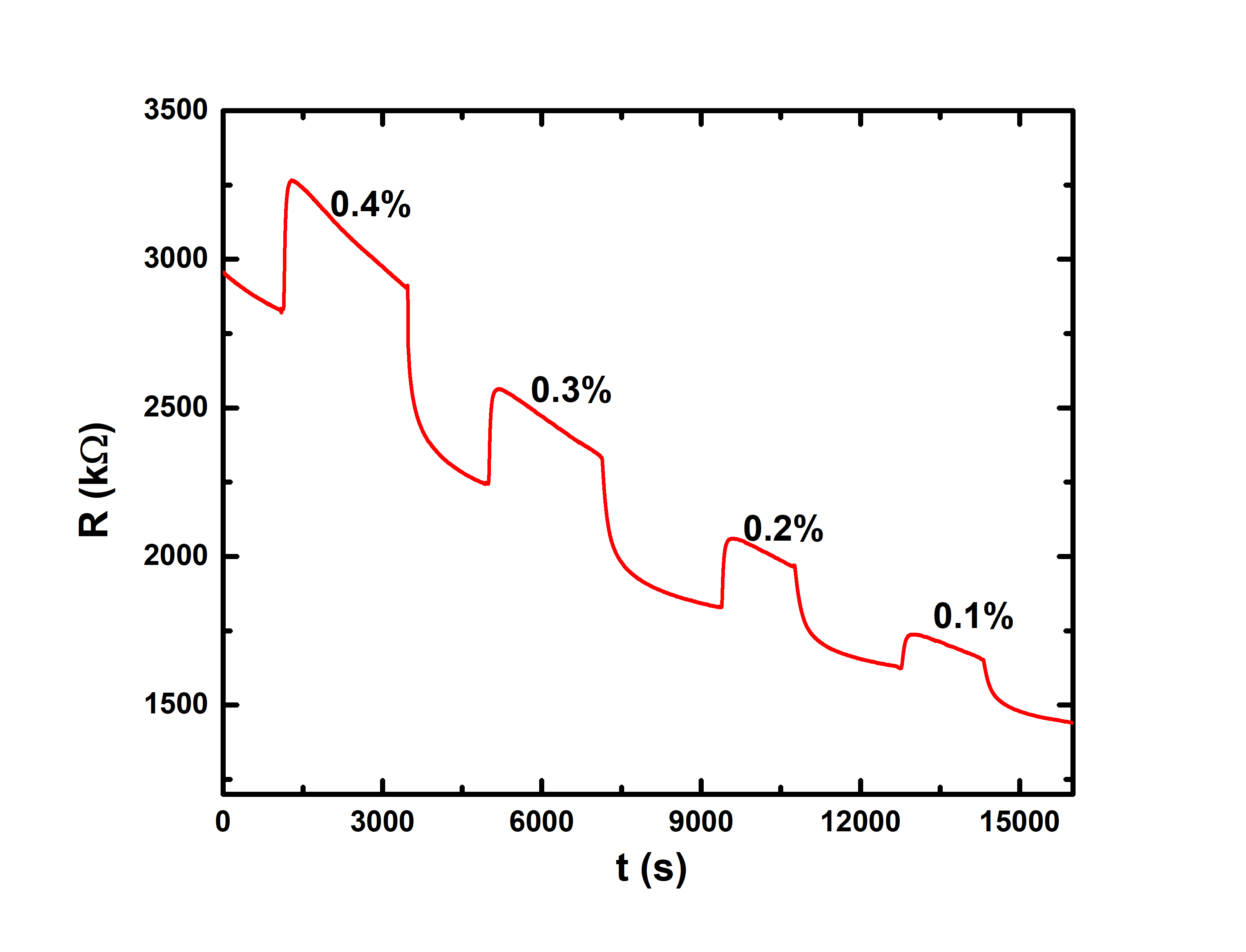}
	\caption{(Colour online): Resistance-time graph for a sensing device with PVP$/$Pd mole ratio of 0.01 M at concentrations of H$_2$ varying from 0.4$\%$ to 0.1$\%$.}
	\label{fig: R-t}
\end{figure}

Our sensing device was found to be quite robust to the amount of the self-assembled monolayer used. The sensing data in the main paper shows a sensing response on a SAM that was made at a concentration of 0.1 M. In order to check if the device is responsive to other concentrations of SAM used, we made an architecture with PVP$/$Pd mole ratio of 0.10 but with an order of magnitude lower concetration of SAM used, namely, 0.01 M. It was found that the architecture was responsive to H$_2$ even at this reduced concentration as shown in Fig. \ref{fig: R-t}. However, the saturation was found to be not as good as that with SAM of 0.1M concentration (see the main paper).  

\section{Comparison of sensing performance of various Pd based resistive sensing devices}
\begin{table}[h!]
	\begin{tabular}{|c |c |c |c |c |c |c |c |}
		\hline
		S. No.  & \makecell{Assembly \\-type}   & \makecell{Fabrication \\technique}  & \makecell{H$_2$ ($\%$)} & \makecell{Gas \\ Conditions} & \makecell{Sensitivity \\($\%$)} & \makecell{Time-constant \\$t_{90}$ (s)} & Ref.\\
		\hline
		1   &    \makecell{Pd-PVP \\architecture} & \makecell{Chemical\\-film}  & 0.5 & H$_2$/Air & 27 & 47 &   \makecell{This \\work}   \\ \hline
		2   &    \makecell{hexanethiolate \\-coated Pd} & \makecell{Chemical\\-film}  & \makecell{0.5 \\1} & H$_2$/N$_2$ & \makecell{1 \\--} & \makecell{-- \\20-50} &   \cite{ibanez}   \\ \hline
		3   &    \makecell{SAM coated \\Pd} & \makecell{e-beam\\evaporation}  & 0.5 & -- & 27 & -- &   \cite{xu}   \\ \hline
		4   &    \makecell{ZnO-InO \\nanorods} & Sputtering  & 0.05 & H$_2$/Air & 20.5 & -- &   \cite{huang}   \\ \hline
		5   &    \makecell{Pd-thin \\films} & Sputtering  & 2 & H$_2$/Air & 5 & -- &   \cite{ramanathan}   \\ \hline
		6   &    \makecell{Pd-thin \\films} & Sputtering  & 1 & H$_2$/N$_2$ & 10 & 130 &   \cite{lee1}   \\ \hline
		7   &    \makecell{Pd-thin \\films} & \makecell{Semiconductor \\ processes}  & 1 & H$_2$/Air & 10 & 40 &   \cite{lee}   \\ \hline
		8   &    \makecell{Pd-thin \\films} & \makecell{Pulsed laser \\ deposition}  & 1 & H$_2$/Air & 7-8 & 10-20 &   \cite{kumar}   \\ \hline
		9   &    \makecell{Pd \\nanowires} & \makecell{Electro \\ -deposition}  & 1 & H$_2$/N$_2$ & 1.3 & 10 &   \cite{zeng}   \\ \hline
		10   &    \makecell{Pd \\nanowires} & \makecell{Electro \\ -deposition}  & 1 & H$_2$/N$_2$ & 0.1 & 700 &   \cite{jeon}   \\ 
		\hline
	\end{tabular}
	\caption{Comparison of the performance of various Pd-based resistive sensors made by different techniques and measured in different gas environments.}
	\label{Table:Comparison}
\end{table}

Table \ref{Table:Comparison} above shows the t$_{90}$ response of various Pd based resistive sensors under different concentrations of H$_2$ gas. From the table, it can be said that our chemiresistive H$_2$ sensor shows excellent sensitivity values and rise time response in ambient H$_2$ conditions, corresponding to a real H$_2$ leak environment.\\

%\bibliography{49}